\documentclass[]{aa}  
\usepackage{graphicx,amsmath,wasysym}
\usepackage{amssymb}
\usepackage{color}
\usepackage{natbib}             
\usepackage{url}
\usepackage{multirow}
\bibpunct{(}{)}{;}{a}{}{,} 
\usepackage{sidecap}

\usepackage{array}
\usepackage{tabularx}

\usepackage{txfonts}
\newcolumntype{Z}{>{\centering\let\newline\\\arraybackslash\hspace{0pt}}X}

\newcommand{\Mearth}{\mathrm{M}_\oplus}
\newcommand{\Rearth}{\mathrm{R}_\oplus}

\begin{document}

    \title{Planetary system around LTT 1445A unveiled by ESPRESSO: Multiple planets in a triple M-dwarf system\thanks{Based on Guaranteed Time Observations collected at the European Southern Observatory under ESO programme(s) 1104.C-0350, 106.21M2.002 by the ESPRESSO Consortium.}}

   \author{
   B.~Lavie\inst{1},
   F.~Bouchy\inst{1},
C.~Lovis\inst{1},
M.~Zapatero Osorio\inst{5},
A.~Deline\inst{1},
S.~Barros\inst{9,11},
P.~Figueira\inst{10},
A.~Sozzetti\inst{4},
J.~I. Gonz\'alez Hern\'andez\inst{6,7},
J.~Lillo-Box\inst{5},
J.~Rodrigues\inst{9,11},
A.~Mehner\inst{10},
M.~Damasso\inst{4},
V.~Adibekyan\inst{9,11},
Y.~Alibert\inst{2},
C.~Allende Prieto\inst{6,7},
S.~Cristiani\inst{3},
V.~D'Odorico\inst{3,12},
P.~Di Marcantonio\inst{3},
D.~Ehrenreich\inst{1},
R.~G\'enova Santos\inst{6,7},
G.~Lo Curto\inst{13},
C.J.A.P.~Martins\inst{11,18},
G.~Micela\inst{17},
P.~Molaro\inst{3,12},
N.~Nunes\inst{8},
E.~Palle\inst{6,7},
F.~Pepe\inst{1},
E.~Poretti\inst{14,15},
R.~Rebolo\inst{6,7,16},
N.~Santos\inst{9,11},
S.~Sousa\inst{11},
A.~Su\'arez Mascare\~no\inst{6,7},
H.~Tabrenero\inst{5},
S.~Udry\inst{1}
}    
\authorrunning{B.~Lavie et al.}
\titlerunning{Planetary system around LTT 1445A unveiled by ESPRESSO}


\institute{
Observatoire de l'Universit\'e de Gen\`eve, 51 chemin Pegasi, 1290 Sauverny, Switzerland \and Physics Institute, University of Bern, Sidlerstrasse 5, 3012 Bern, Switzerland \and INAF - Osservatorio Astronomico di Trieste, via G. B. Tiepolo 11, I-34143, Trieste, Italy \and INAF - Osservatorio Astrofisico di Torino, via Osservatorio 20, 10025 Pino Torinese, Italy \and Centro de Astrobiolog\'ia (CSIC-INTA), Crta. Ajalvir km 4, E-28850 Torrej\'on de Ardoz, Madrid, Spain \and Instituto de Astrof\'isica de Canarias (IAC), Calle V\'ia L\'actea s/n, E-38205 La Laguna, Tenerife, Spain \and Departamento de Astrof\'isica, Universidad de La Laguna (ULL), E-38206 La Laguna, Tenerife, Spain \and Instituto de Astrof\'isica e Ciencias do Espaco, Faculdade de Ciencias da Universidade de Lisboa, Edif\'icio C8, Campo Grande, PT1749-016 Lisbon, Portugal \and Departamento de F\'isica e Astronomia, Faculdade de Ciencias, Universidade do Porto, Rua Campo Alegre, 4169-007, Porto,Portugal  \and European Southern Observatory, Alonso de Cordova 3107, Vitacura, Region Metropolitana, Chile \and Instituto de Astrof\'isica e Ciencias do Espaco, CAUP, Universidade do Porto, Rua das Estrelas, 4150-762, Porto, Portuga \and Institute for Fundamental Physics of the Universe, Via Beirut 2, I-34151 Miramare, Trieste, Italy \and European Southern Observatory, Karl-Schwarzschild-Strasse 2, 85748, Garching b. Munchen, Germany \and  Fundaci\'on Galileo Galilei-INAF, Rambla Jos\'e Ana Fernandez P\'erez 7, 38712 Bre\~na Baja, TF, Spain \and INAF - Osservatorio Astronomico di Brera, Via E. Bianchi 46, I-23807 Merate, Italy \and Consejo Superior de Investigaciones Cient\'icas, Spain \and INAF - Osservatorio Astronomico di Palermo, Piazza del Parlamento 1, I-90134 Palermo, Italy \and Centro de Astrof\'{\i}sica da Universidade do Porto, Rua das Estrelas, 4150-762 Porto, Portugal
}
   
   \date{} 

  \abstract{We present radial velocity follow-up obtained with ESPRESSO of the M-type star LTT 1445A (TOI-455), for which a transiting planet b with an orbital period of~5.4 days was detected by TESS. We report the discovery of a second transiting planet (LTT 1445A c) and a third non-transiting candidate planet (LTT 1445A d) with orbital periods of 3.12 and 24.30 days, respectively. The host star is the main component of a triple M-dwarf system at a distance of 6.9 pc. We used 84 ESPRESSO high-resolution spectra to determine accurate masses of 2.3$\pm$0.3 $\Mearth$ and 1.0$\pm$0.2 $\Mearth$ for planets b and c and a minimum mass of 2.7$\pm$0.7  $\Mearth$  for planet d.    
  Based on its radius of 1.43$\pm0.09$ $\Rearth$ as derived from the TESS observations, LTT 1445A b has a lower density than the Earth and may therefore hold a sizeable atmosphere, which makes it a prime target for the James Webb Space Telescope. We used a Bayesian inference approach with the nested sampling algorithm and a set of models to test the robustness of the retrieved physical values of the system. There is a probability of 85$\%$ that the transit of planet c is grazing, which results in a retrieved radius with large uncertainties at 1.60$^{+0.67}_{-0.34}$ $\Rearth$. LTT 1445A d orbits the inner boundary of the habitable zone of its host star and could be a prime target for the James Webb Space Telescope.
}

\keywords{planetary systems - stars: individual (TOI-455; LTT 1445) - techniques: radial velocities; photometric}

  \maketitle

\section{Introduction} 

\textit{The Transiting Exoplanet Survey Satellite} (TESS) NASA mission \citep{Ricker2018} is a powerful observatory for detecting potential transiting candidates. To date, it has provided more than 4000 candidates. \textit{The Echelle Spectrograph for Rocky Exoplanet and Stable Spectroscopic Observations}, ESPRESSO \citep{Pepe2021}, operating at the Very Large Telecope (VLT) in Chile, serves the community as an efficient tool for following these candidates up and allows determining radial velocity (RV) measurements with a 10 cm.s$^{-1}$ precision. Because radial velocity observations are limited by photon noise precision and instrumental instability, it is natural to use ESPRESSO to determine the masses of small-size planets with a radius smaller than two Earth radii ($\Rearth$). The instrument is indeed mounted on the VLT, and its initial design was oriented towards high stability. This is necessary requirement for reaching the precision that is required to detect these planets, which is one of the objective of the ESPRESSO Guaranteed Time Observation (GTO) consortium (Program ID 1102.C-744, PI: F.Pepe). The subprogram of the ESPRESSO GTO that focuses on determining precise masses of TESS and K2 transiting candidates has produced several results since its start on October 1, 2018: \cite{Damasso_2020} derived a precise characterisation of the $\Pi$ Mensae system, which is composed of a brown dwarf and a 4.3 $\pm$0.7 Earth-mass ($\Mearth$) super-Earth; the masses of the super-Earth and sub-Neptune planets around the G2 star K2-38 were determined by \cite{Toledo_2020} as 7.3 $_{-1.0}^{+1.1}$$\Mearth$ and 8.3$\pm$1.3 $\Mearth$, respectively; \cite{Sozzetti_2021} unveiled two planets around the bright late-F dwarf HD 5278, a 7.8$_{-1.4}^{+1.5}$$\Mearth$ transiting sub-Neptune and a non-transiting Neptune at periods of 14 and 41 days, respectively;  \cite{Mortier2020} reported two planets in near-resonance around the slightly evolved G2 star K2-111 at 5.3 and 15.7 days with a mass of 5.29$_{-0.77}^{+0.76}$$\Mearth$ for the inner planet that transits in front of its host star; and finally, \cite{Demangeon_2021} announced the detection of the lowest-mass planet measured so far around L 98-59, a nearby stellar system with four planets.

The determination of the density for planets with measured radius and mass allows inferring their bulk composition. Current observations suggest that exoplanets follow two distinct trends: similar to an Earth-like composition for planets with a mass of up to $\sim$25 $\Mearth$ and a more volatile-rich composition for planets more massive than $\sim$10 $\Mearth$ \citep{Otegi_2020} . The transition of these two populations and their origins are key questions in exoplanet research and require a precise mass determination for both populations. Using an updated exoplanet catalogue based on reliable and robust masses, \cite{Otegi_2020} showed that the dispersion around the Earth-like composition for the rocky exoplanet population is small. Theoretical models show inherent degeneracies in the composition of planets with similar masses and radii, especially in the range of Earth masses to Neptune masses (e.g. \citealt{Sotin_2007,Seager_2007,Howe_2014,Dorn_2017}); however, the lack of precise measurements of the two physical variables remains the dominant source of uncertainty in this analysis. 
Efforts to increase the number of planets with accurate mass and radius measurements are therefore paramount. In the near future, the need for a broad sample of planets with a high-accuracy density will be necessary to identify suitable targets for an atmospheric characterisation with the next generation of observatories, such as the James Webb Space Telescope  (JWST) \citep{Gardner_2006,Deming_2009} or the Extremely Large Telescope (ELT) \citep{Gilmozzi_2007,Marconi_2016,Marconi_2020,Marconi_2021}, and also to serve as input in the atmospheric models that will study these targets. 

M dwarfs account for roughly 75$\%$ of all the stars in our Galaxy \citep{Henry_2006}. They provide stronger radial velocity signals and a high contrast ratio
and transit depth for transiting Earth-like planets that favours the detection of spectral features in the atmosphere of these planets \citep{Wunderlich_2019}.
The probability of hosting multiple planets in compact orbits is higher for M dwarfs than other stars \citep{Muirhead_2015,Ballard_2019}. Hence planets transiting M dwarfs are prime targets in our quest for accurate mass and radius measurements, based on which planets with an atmosphere can be identified. We also wish to be able to study planet composition as a function of stellar properties, and it is therefore required to first detect and characterise planets around stars with lower masses.

Stellar multiplicity is a decreasing function of the primary mass \citep{Duchene_2013}. The multiplicity rate, that is, the percentage of stellar systems with more than one star, is therefore higher for solar-type stars (44$\%;$  \citealt{Duquennoy_1991,Raghavan_2010,Tokovinin_2014}) than for M dwarfs (27 $\%;$  \citealt{Leinert_1997,Duchene_2013,Ward_2015,Winters_2019}). For these two stellar types, binaries form the bulk of the contingent and triple star systems account for 8$\%$ and 3$\%$ of the overall population of solar-type stars and M dwarfs, respectively. The study of planets within these systems provides a unique angle to help shape our global understanding of the planet formation mechanisms. The required stability over a period of time of planetary systems in binary or higher-order systems that is long enough adds complexity to the problem of habitability around M dwarfs \citep{Shields_2016} and forces a deeper determination of the very notion of habitability.

The LTT 1445 ABC system is a triple stellar system composed of three M dwarfs in a hierarchical configuration. A compact binary orbits the main component at a minimum separation of $\sim$ 21 astronomical units (AU). A TESS candidate has been confirmed by \cite{Winters2019} with only one TESS sector. The system is at a distance of only 6.9 parsecs (pc) from Earth, which makes it the second closest transiting exoplanet system to date after HD~219134 \citep{motalebi_2015}. The planet orbits the main component LTT 1445A with an orbital period of $\sim$ 5.4 days and has a radius of $\sim$1.4$\Rearth$, which makes it an ideal target for ESPRESSO follow-up. 

In this paper, we present radial velocity measurements obtained with ESPRESSO of the M dwarf LTT 1445A (TOI-455) that allow us to determine the mass for the previously announced transiting planet b with an orbital period of 5.4 days with high accuracy. We also report a new transiting planet c at 3.12 days and a potential non-transiting planet d at 24.30 days.
Section 2 presents the observations and Sect. 3 the stellar properties. In section \ref{Photo_Tess} we detail the photometric analysis of the TESS observations, including an additional sector with respect to  \citet{Winters2019}. In section \ref{RVA} we present the radial velocity analysis with an investigation of the stellar activity of the host star. The last section reports the results from the combined photometric and spectroscopic analysis and a final discussion.

\section{Observations} \label{sec:obs}
\subsection{TESS photometry}
The LTT 1445 system has been observed by TESS in sector 4 (from 19 October to 14 November 2018; a 26-day interval) and sector 31 (from 22 October to 18 November 2020; a 27-day interval). Both sectors were observed with CCD 4 of Camera 2 with a 2-minute cadence. The 20-second cadence mode is also available for sector 31. Following the method of \citet{LilloBox2020}, we used tpfplotter \footnote{\label{tpfplotter}https://github.com/jlillo/tpfplotter} \citep{Aller_2020} to scan for any contaminant sources. As shown in Figure \ref{tpfplotter}, we did not find any contaminants with magnitude contrast brighter than 8 in the automatically selected aperture.
\begin{figure}
\centering
\includegraphics[width=\columnwidth]{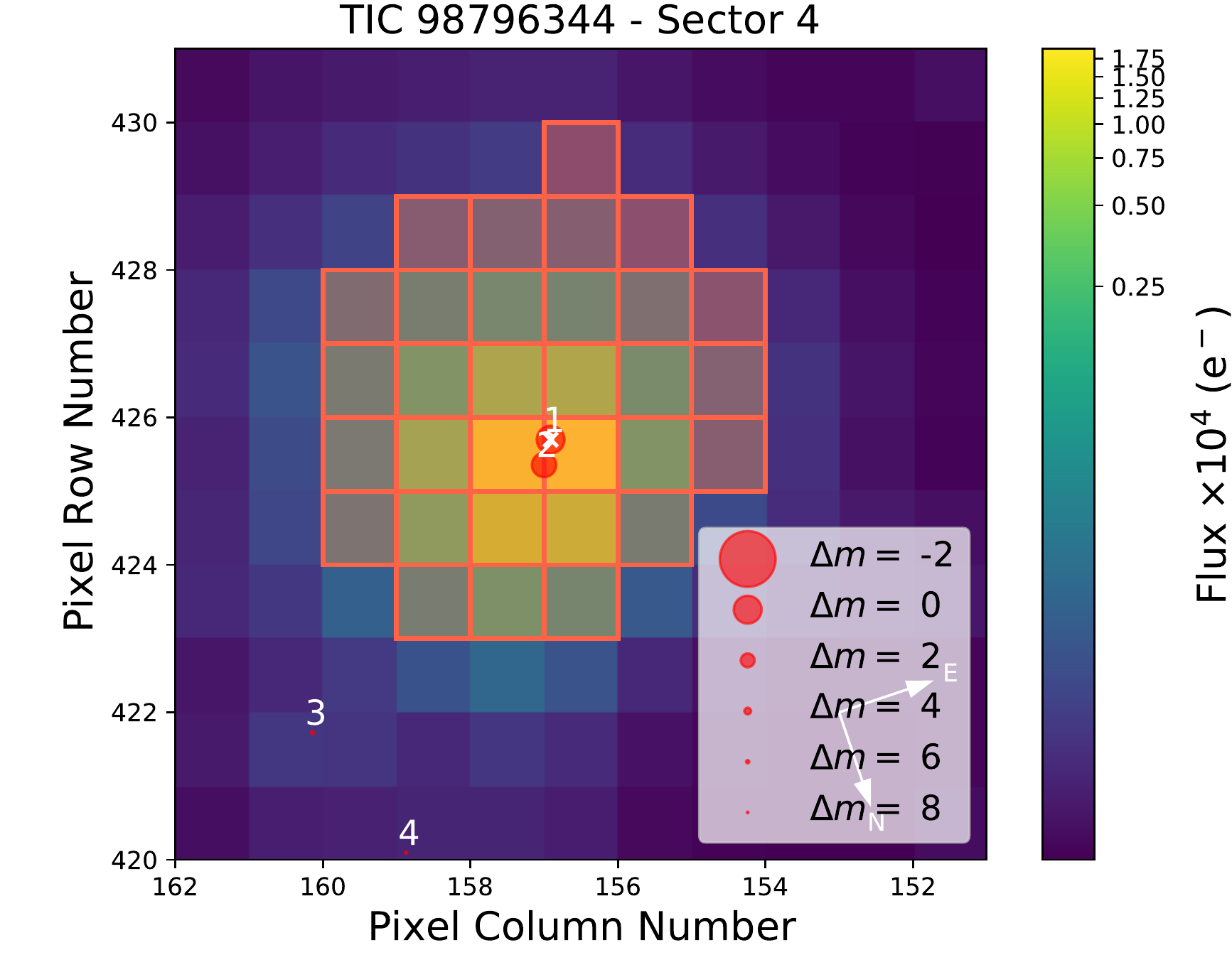}
\caption{\label{tpfplotter} TPF of LTT 1445 from the TESS observations in Sector 4
(composed with tpfplotter; \citealt{Aller_2020}). The SPOC pipeline
aperture is overplotted with shaded red squares, and the Gaia DR3 catalogue is also overlaid with symbol sizes proportional to the magnitude
contrast with the target, marked with a white cross. We display all sources from the Gaia DR3 catalogue with a magnitude contrast up to $\Delta$m = 8 mag.}
\end{figure}

\cite{Winters2019} detected a transiting planet with an orbital period of $\sim$5.4 days using sector 4 around LTT 1445A. 
The publicly available TESS photometric data were downloaded from the Mikulski Archive for Space Telescopes (MAST).

\subsection{Spectroscopic follow-up with ESPRESSO}
We carried out the radial velocity monitoring of the system with the ESPRESSO spectrograph installed at the VLT in the Paranal Observatory in Chile. The observations were part of the subprogram of the GTO that focuses on determining precise masses of TESS and K2 transiting candidates. All observations were carried out with the 1UT mode, no binning, and 900-second exposure time with the calibration source (fiber B) pointed to the Fabry Perot. The spectroscopic data were extracted using version 2.3.0 of the ESPRESSO data reduction pipeline (Lovis et al. in prep.), which is available for direct download from the ESO pipelines website\footnote{\label{esosoft}http://www.eso.org/sci/software/pipelines/}. The data reduction was performed with an M4 mask. The extracted data are downloaded from the DACE platform API\footnote{\label{dace}The Data Analysis Center for Exoplanets (DACE) platform is available at https://dace.unige.ch}.
The ESPRESSO dataset consists of 84 observations taken between 20 July 2019 and 12 March 2021 (the observation with ESO OB Id 2549082 was stopped after $\sim$500 seconds of exposure time and is not considered in this dataset). The gap in the time coverage of these observations is due to the Covid crisis, which caused the instrument at the Paranal observatory to be unavailable for nearly nine months.
The ESPRESSO dataset is combined with 14 observations that are available in the ESO HARPS archive (Program ID 072.C-0488, PI: M.Mayor; Program IDs 183.C-0437 and 1102.C-0339, PI: X.Bonfils) that were also used in \cite{Winters2019}.  
The median uncertainty for the radial velocity time-series is 0.48 m.s$^{-1}$ and 1.16 m.s$^{-1}$ for the ESPRESSO and HARPS datasets, respectively.
The combined dataset (Figure \ref{rv_all}) shows a clear long-term drift due to the interaction of LTT 1445A with the BC component of the triple system.

\section{Stellar properties}\label{stellarprop}
\subsection{Spectral energy distribution}
\begin{figure}
\centering
\includegraphics[width=\columnwidth]{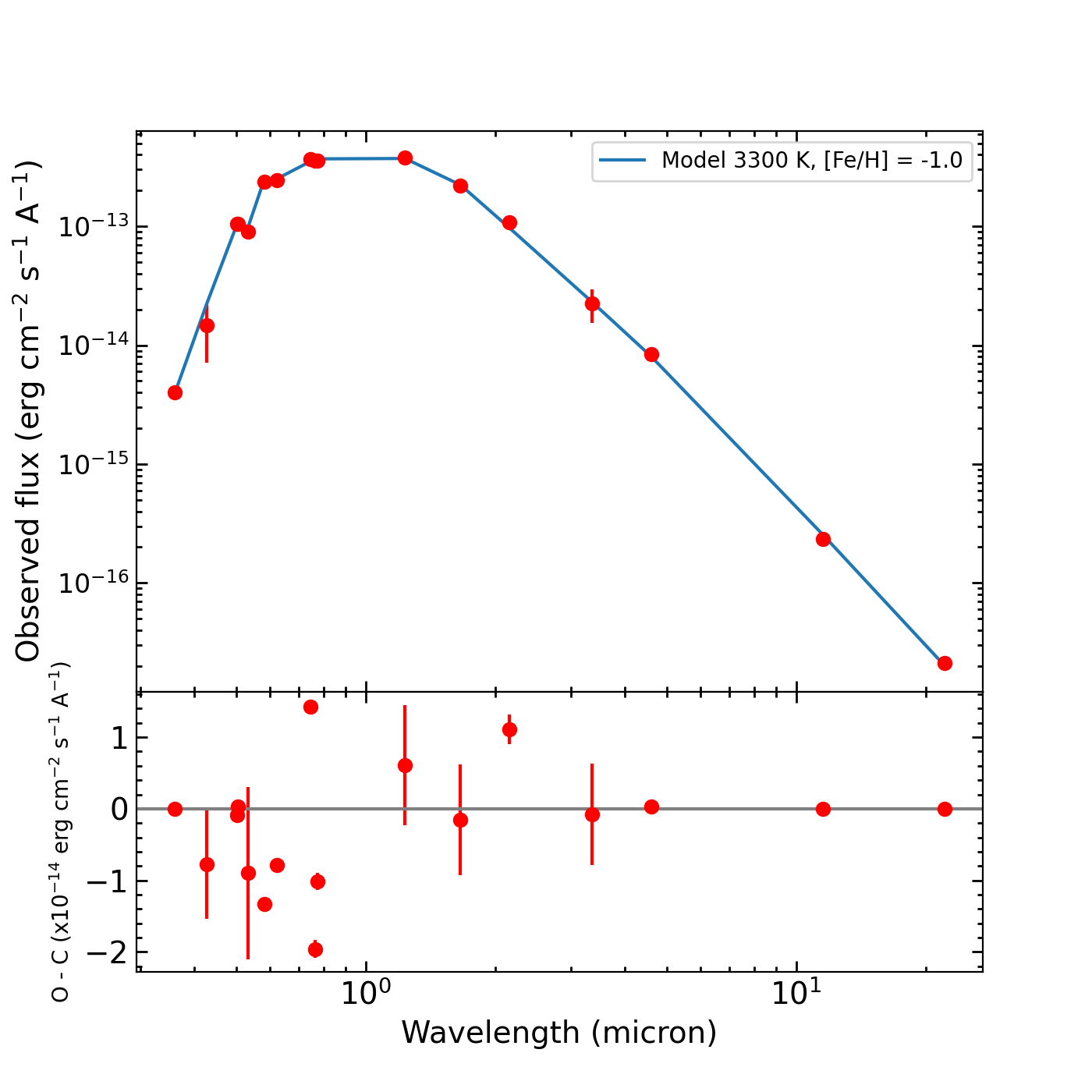}
\caption{Observed photometric spectral energy distribution of LTT 1445A (red dots) from the $u$ filter through the $W4$ band. The best BT-Settl model \citep{2012RSPTA.370.2765A}, corresponding to $T_{\rm eff}$ = 3300 K, low metallicity, and log\,$g$ = 6.0 (cgs), is shown with the blue line.  \label{fig:sed}}
\end{figure}

We estimated the physical parameters of LTT 1445A by fitting the photometric spectral energy distribution constructed from catalogue broadband magnitudes following the prescription of the Virtual Observatory VOSA tool \citep{2008A&A...492..277B,Rodrigo_2020}. The available photometry includes optical magnitudes from {\sl Gaia}, APASS, Tycho-2, and the Sloan Digital Sky Survey catalogues \citep{2015AAS...22533616H, 2000A&A...355L..27H, 2017AJ....154...28B}, the $JHK$ magnitudes from the Two Micron All Sky Survey (2MASS; \citealt{2006AJ....131.1163S}), and the $W1-4$ magnitudes from the Wide-field Infrared Survey Explorer mission (AllWISE; \citealt{2010AJ....140.1868W, 2011ApJ...731...53M}). We employed the BT-Settl models \citep{2012RSPTA.370.2765A}, which are computed for a wide range of effective temperatures ($T_{\rm eff}$), surface gravities (log\,$g$), and atmospheric metallicities ([Fe/H]) with steps of 100 K, 0.5 dex, and 0.5 dex, respectively. The photometric spectral energy distribution of LTT 1445A  covers the wavelength interval 0.36--22.09 $\mu$m. We assumed no interstellar extinction due to the proximity of the star ($d$ = 6.864 $\pm$ 0.001 pc) and no obvious reddening in the stellar colours at the long wavelengths of AllWISE data. From this analysis, we obtained $T_{\rm eff}$ = 3300 $\pm$ 100 K, log\,$g$ = 6.0 $\pm$ 0.50 dex, and [Fe/H] = $-1$ dex. The high surface gravity is consistent with the star being older than $\sim$1 Gyr. The observed spectral energy distribution of LTT 1445A and its best fit are shown in Fig.~\ref{fig:sed}. By integrating the observed spectral energy distribution, we derived the stellar luminosity to be log\,$L/L_\odot$ = $-2.099 \pm 0.010$ dex.

\begin{figure}
\centering
\includegraphics[width=\columnwidth]{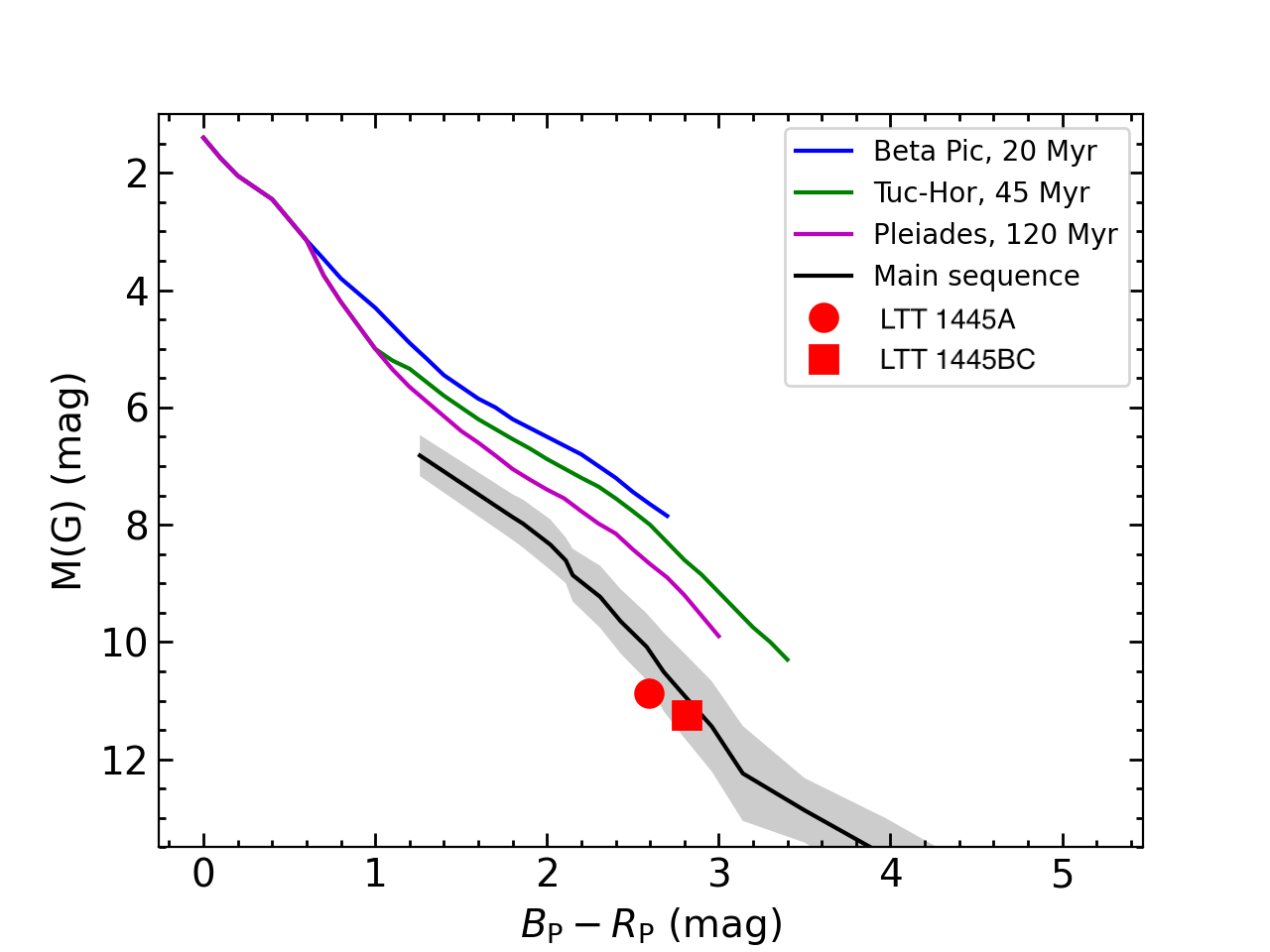}
\caption{Location of LTT 1445A and BC in the {\sl Gaia} colour-magnitude diagram. The binary BC is not resolved by {\sl Gaia}. The young stellar isochrones of the $\beta$ Pic (blue) and Tucana-Horologium (green) moving groups and the Pleiades cluster (purple) are taken from \citet{2018AJ....156..271L}, and the main sequence of field stars (black) is taken from \citet{2020A&A...642A.115C}. The grey area represents the dispersion observed among main-sequence stars. \label{fig_hr}}
\end{figure}

\subsection{Mass and radius of LTT 1445A}
The location of LTT 1445A in the {\sl Gaia} colour-magnitude diagram is displayed in Figure~\ref{fig_hr}, where the position of the fainter binary LTT 1445BC is also indicated for completeness.  The magnitudes and colours of LTT 1445A are consistent with M4.25 $\pm$ 0.25. All stellar sequences shown in the figure were built using {\sl Gaia} photometry and parallaxes (see \citealt{2018AJ....156..271L} for the young isochrones and \citealt{2020A&A...642A.115C} for the main-sequence track), thus minimising any possible systematics in the comparison exercise. The absolute $G$-band magnitude of LTT 1445A is fainter than that of the main sequence for its colour, which is consistent with a slightly sub-solar metallicity, as was inferred from the analysis of the stellar photometric spectral energy distribution. 
The position of LTT 1445BC appears to lie closer to the main sequence of stars, but the two components of the subarcsecond binary were not resolved by {\sl Gaia}. When the photometry was deblended into individual magnitudes following the delta-magnitudes provided by \citet{Winters2019}, the positions of LTT 1445B and C lie at fainter magnitudes in Figure~\ref{fig_hr} and away from the solar-metallicity track. The colours and absolute magnitudes of the stars are compatible with their being main-sequence M-type stars.

At low stellar masses, results derived from stellar models (either evolutionary or atmosphere models) strongly depend on the assumed physics, which are not perfectly controlled at low temperatures. To avoid this problem as much as we can, we therefore decided to base our mass and radius deteminations of LTT 1445A on relations constructed from compilations of detached double-lined double-eclipsing main-sequence M-dwarf binaries \citep{2019ApJ...871...63M}.
Using the most recent {\sl Gaia} distance, the 2MASS $K$ magnitude, and the empirical mass--$M_K$ \citep{2019ApJ...871...63M} and radius--$M_K$ \citep{2015ApJ...804...64M} relations, we determined the mass and size of LTT 1445A as $M = 0.249 \pm 0.023$ M$_\odot$ and $R = 0.276 \pm 0.010$ R$_\odot$ for solar metallicity. For a sub-solar metallicity of [Fe/H] = $-1.0$ dex, the radius would slightly increase to $R = 0.289 \pm 0.010$ R$_\odot$ , while the stellar mass remains unchanged within the quoted uncertainty for metallicities around solar, as discussed in \citet{2019ApJ...871...63M}. The errors in mass and radius were derived from the uncertainties in distance, $K$-band magnitude, and the dispersion of the empirical relations. 
These values are compatible with those of \citet{Winters2019}.

\section{Photometry analysis with TESS}\label{Photo_Tess}
\subsection{Identification of transit candidates}\label{identification_transit}
The three stars of the system LTT 1445 are included in the TESS aperture because of the large TESS pixel size of 21". We used the \texttt{lightkurve} python package \citep{lightkurve2018} to download the three light curves produced by the Presearch Data Conditioning Simple Aperture Photometry  (PDCSAP; \citealt{Smith2012,Stumpe2012,Stumpe2014}) algorithm for both sectors. The three light curves for the two sectors are relatively complex and exhibit a clear modulation signal at 1.4 days (Figure \ref{detrendedlc}). 
In this section, we perform a blind initial analysis to identify potential planet candidates.

\subsubsection{Modulation signal at 1.4 days}\label{signal1pt4}

The modulation signal with a period of 1.4 days is clearly visible in both sectors. \cite{Winters2019} suspected that it originates from either the B or C component, based on the activity indicator measurements from spectra obtained with the Tillinghast Reflector Echelle Spectrograph. The amplitude of the signal varies between the two epochs with a root mean square (rms) of 2.37 and 4.13 part per thousand (ppt) during observation of sector 4 and sector 31, respectively.

Similarly to \cite{Winters2019}, we chose to model it as well as any other stellar signals with a Gaussian process (GP) \citep{RasmussenWilliams2006}. A GP is characterised by a covariance matrix generated by a covariance function (kernel) that determines the characteristics of the underlying model. The choice of kernel and its hyperparameters is arbitrary in the sense that it is not based on some physical first principles. However, it is conditioned by our understanding of the underlying physics (periodicities and timescales of the expected signals). Throughout this paper, we use a kernel with three hyperparameters: the amplitude B, the length L, and the period P.
An arbitrary element, $C_{ij}$, of the covariance matrix is
\begin{equation}
C_{ij} = \frac{B}{2} * \exp( -\frac{|t_i - t_j|}{ L}) * (\cos(\frac{ 2\pi|t_i-t_j| }{ P}) + 1 )
.\end{equation}
This kernel is a simpler version of the kernel described in \cite{celerite}, which has an additional scale parameter C. Models using parameter C were systematically rejected when compared with models without it. We therefore did not justify the use of this additional parameter. 
This kernel was implemented with the \textit{celerite} package \citep{celerite}.

The three PDCSAP light curves were normalised, time windows affected by flares (spotted by eye) were removed, and a 3-$\sigma$ clipping was performed on the time series to remove outliers. The light curves were corrected for dilution, and we kept the dilution factor provided by the pipeline in the keyword CROWDSAP. The light curves were binned with 500 bins. All these operations were made with the \texttt{lightkurve} package. We then performed a Bayesian inference on the hyperparameters of the GP with the three binned light curves independently. The results are similar for the two cadences of sector 31. We also performed the inference with the two sectors combined using the 2-minute cadence only. The inference was made with the nested sampling algorithm \citep{skilling_2006} implemented with \texttt{pymultinest} \citep{buchner_x-ray_2014}, a Python wrapper for \texttt{multinest} \citep{feroz_importance_2013,feroz_multinest:_2009,feroz_multimodal_2008}. The nested sampling algorithm computes the Bayesian evidence (the logarithm of the Bayesian evidence is noted ln $\mathcal{Z}$), which allows comparing the models, but also provides the posterior distribution of the parameters as a byproduct. The three hyperparameters converged to similar values for the four inferences (see Appendix Fig \ref{post_hyperparam_1pt4}).

\subsubsection{Identification of two transit signals}\label{twotransitsignal}
\begin{figure*}
\centering
\includegraphics[width=2\columnwidth]{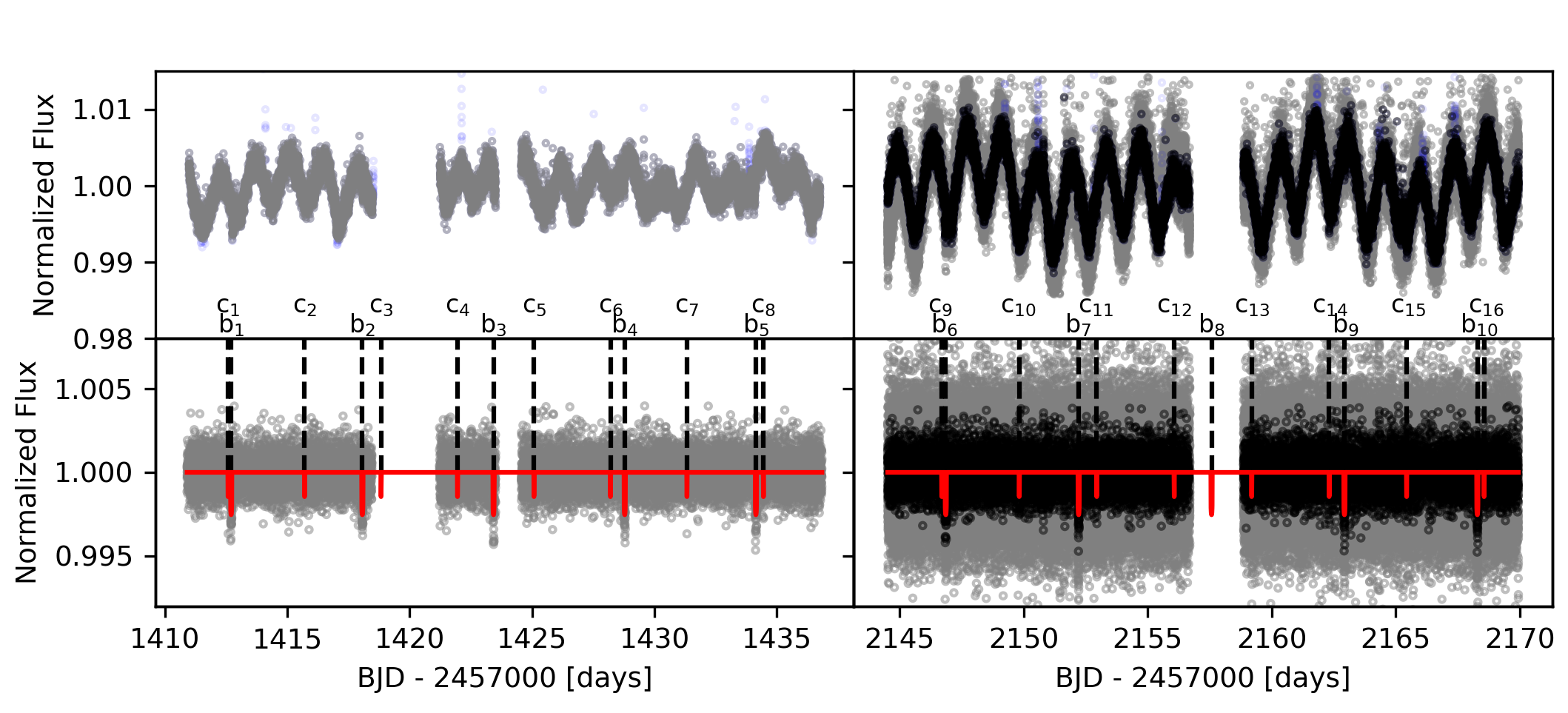}
\caption{\label{detrendedlc} LTT1445 ligt curves from TESS sectors 4 and 31. First row: TESS PDCSAP light curve for sector 4 with 2-minute cadence (first column) and sector 31 (second column) for a 2-minute cadence (black) and a 20-second cadence (grey). A clear modulation with a period of~1.4 days is  clearly visible. Second row: Respective detrended light curves with the best-fit model (in red) from the analysis of section \ref{Transit_res}. The vertical lines identify the transit for each planet: b$_1$ to b$_{10}$ for planet b, and c$_1$ to c$_{16}$ for planet c. The light blue dots in the 2-minute cadence in the upper panels are the outliers in the dataset that were removed from the  PDCSAP light curves. }
\end{figure*}

The light curves were detrended using the mean prediction of the GP with the hyperparameters set to the maximum a posteriori (MAP) values obtained through the Bayesian inference on the combined dataset (2-minute cadence for both sectors).
We performed a periodogram analysis to identify potential transits within each light curve using the box least-squares (BLS) method for periods between 1 and 20 days and durations between 0.01 and 0.5 days. When a transit signal was found, it was masked to allow another iteration of the periodogram analysis. Results are shown in Appendix \ref{iterative_period_lc}.
The strongest detected transit signal is linked to the planet that orbits LTT 1445A at 5.4 days \citep{Winters2019}. A second signal is clearly detected in the two sectors at 3.1 days; it might originate from any of the three stars in the system. The light curves stripped of these two signals do not exhibit any significant signals. Fixing the hyperparameters to the MAP values obtained with the independent inference on each light curves does not affect the identification of these two signals.

The transit signals are indicated by b$_i$ and c$_i$ in Figure \ref{detrendedlc}. During the time window of the dataset for the 5.4-day signal, 9 transits took place. One transit (b$_8$) falls in a gap in the light curve. Transit b$_3$ is at the edge of a gap in sector 4 and was not included in \citet{Winters2019}. During the time window of the dataset for the 3.1-day signal, 16 transits occurred. One transit (c$_3$) falls in a gap in the sector 4 light curve. The transit depths are about 0.23$\%$ and 0.16$\%$ for the 5.4- and 3.1-day signals, respectively.

\subsection{TESS transit analysis}

In this section, we perform a Bayesian inference with the 2-minute cadence PDCSAP light curves of both sectors. The light curves were cleaned in a procedure similar to that described in section \ref{signal1pt4}, but without any binning (hereafter, the cleaned light curves). We assumed that the potential planet c orbits the same star as LTT 1445A b, which is confirmed by the radial velocities analysis.

\subsubsection{Joint GP-transit model}
Our model includes a limb-darkening transit model implemented with the Python package \texttt{Batman} \citep{batman2015} and a GP with the kernel from section \ref{signal1pt4}. The hyperparameters of the GP were fixed to the value obtained by the inference on the combined binned light curves. Each planet was modelled with at least five free parameters: the time of inferior conjunction T$_0$, the orbital period P, the planet radius in stellar radii R$_p$/R$_*$, the semi-major axis in stellar radii a/R$_*$, and the orbital inclination i for circular orbits. In case of eccentric orbits, planets were modelled with seven free parameters (the eccentricity $e$ and the longitude of periastron $\omega$ are added to the five retrieved parameters described above). When the model included two or more planets, we enforced Kepler's third law by reducing the parameter space of the additional planets by one parameter. Assuming that the masses of the planet are negligible compared to the mass of the star, we computed the semi-major axis from the period of the planet and the semi-major axis and orbital period of the first planet in the model. We used a quadratic limb-darkening law with two parameters u$_1$ and u$_2$. The parameter space was explored using the nested sampling algorithm. We combined at least three nested sampling runs with 1000 live points and a sampling efficiency of 0.8 to ensure a thorough exploration of the parameter space.
The priors (similar to those in Table \ref{SumTable})  were chosen uniform and broad, except for the periods and the time of inferior conjunction, which were uniform and centred on the values obtained in the quick analysis of section \ref{twotransitsignal} with a width broad enough to capture any shift of their values compared to the quick analysis.

\subsubsection{Transit model selection}\label{Transit_res}

The Bayesian inference was first performed simultaneously with the 2-minute cadence light curves. The Bayesian evidence computed by the nested sampling algorithm allows a comparison of different models with different numbers of parameters (see \citealt{Trotta:2008aa} for insights on Bayesian inference and model selection using the Bayes factor). We tested four models to evaluate the robustness of the two transiting planets: a simple line (with one parameter: a constant), a model with only planet b (7 parameters), a model with planets b and c on circular orbits (11 parameters), and a model with planets b and c on eccentric orbits (15 parameters). The two-planet models are highly favoured, with a logarithm of the Bayes factor (ratio of the Bayesian evidence) higher than 100 between them and the two other models. The logarithm of the Bayes factor between the eccentric and the circular orbit models is only 1.7, which indicates a weak preference for the eccentric orbits because photometric observations only provide weak constraints on the orbital eccentricity. The two additional parameters ( e and $\omega$) improve the fit to the data, but not sufficiently so to compensate for the increase in parameter space volume, which results in a slight improvement of the Bayesian evidence. 
In a second step, we performed a Bayesian inference in which planets b and c were separately modelled with circular orbits on each light curve to probe any differences between the retrieved values of the planet parameters for the two sectors. We tested both cadences for sector 31, which yielded similar results within one sigma. Hence, we only used the results for the 2-minute cadence of both sectors for the rest of our analysis.

\subsubsection{Monitoring the joint GP-transit model: A linear model}
In order to monitor the performance of the joint GP-transit model, we also used a more simple approach with a simple linear model and a \textit{\textup{reduced dataset}} covering the light curves around each transit only. Using the ephemeris obtained through the Bayesian inference with the joint GP-transit model, we identified each transit of each planet in the cleaned light curves. The data were cut at two hours before and after each mid-transit time. We then hid the transit and performed a simple fit with a linear model on the narrow dataset. The data were then detrended with the best-fit linear model. For each sector, we performed a Bayesian inference with a circular transit model for the two planets on the detrended data. Figures \ref{gp_vs_poly_fit_b} and \ref{gp_vs_poly_fit_c} show the \textit{\textup{reduced dataset}} with the best fit from the Bayesian inference with the linear model and the joint GP-transit model (results from the previous section, the inference was made on the full light curves). Both models provide similar fits on the \textit{\textup{reduced dataset,}} showing that the GP component is almost linear around the transits, except for transit b3, which occurs in a curved part of the 1.4-day modulation signal. A cut in the data shorter than two hours or a two-degree polynomial model did not significantly change the inference for the two planets.



\subsubsection{Sensitivity of T$_0$ and TTV}\label{GP_TTV}
\begin{figure}
\centering
\includegraphics[width=\columnwidth]{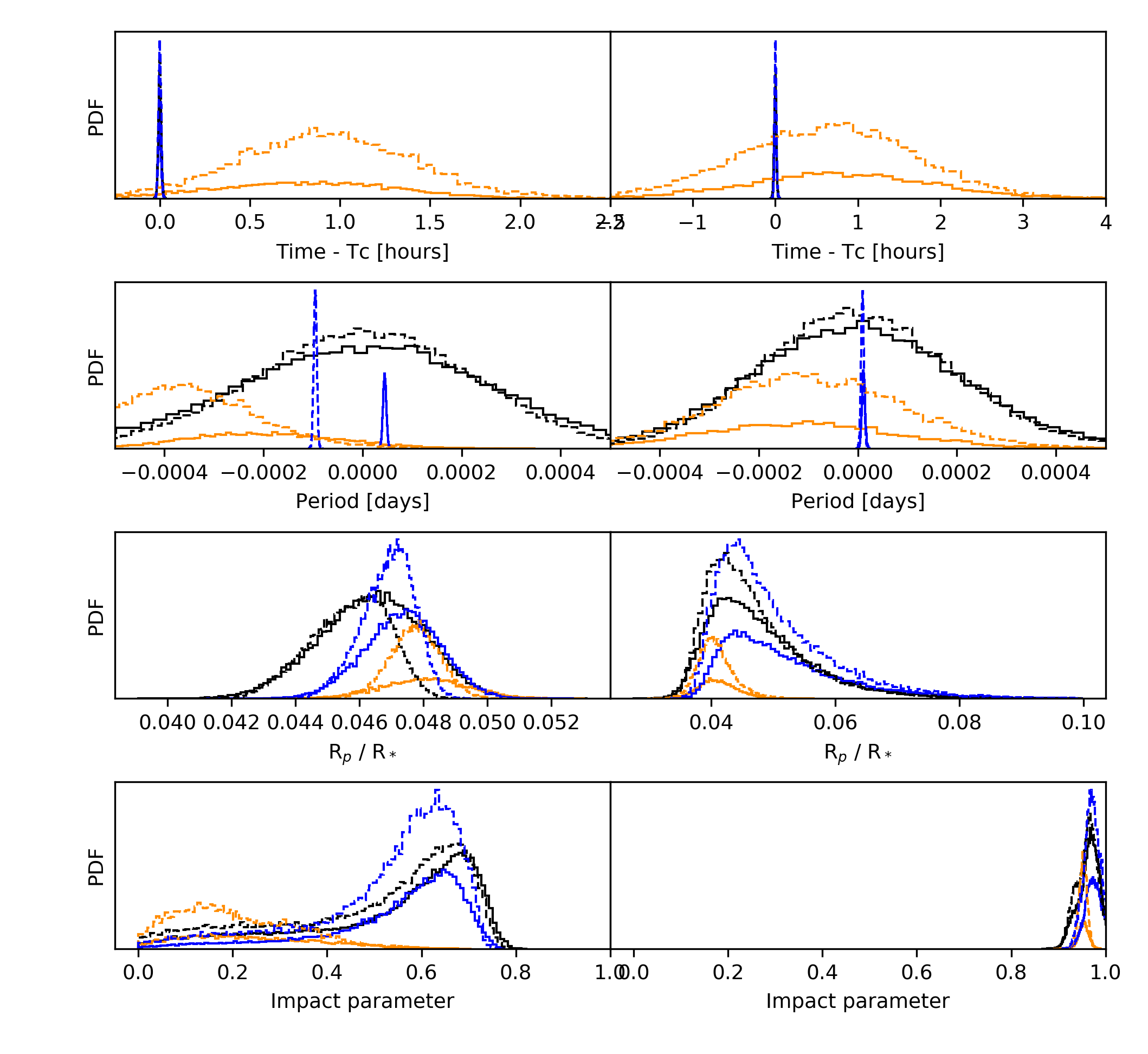}
\caption{\label{Fig_ttv} Posterior distributions for the Bayesian inference with the joint GP-transit model (plain lines) and linear model (dashed lines) for the sector 4 light curve (black), for  sector 31 with the 2-minute cadence (orange), and on the combined 2-minute cadence light-curve dataset (blue). From top to bottom: Time of inferior conjunction, period, ratio of the planet to star radius and impact parameter for planet b (first column) and planet c (second column). Inference on sector 31 shows a broad distribution and a one-hour time transit variation and shift in the orbital period. This broad distribution seems to result from the higher amplitude of the polluting signal at 1.4 days for sector 31. The inference on the combined light curve dataset converges towards the same values as the inference on sector 4. 
 }
\end{figure}

The independent Bayesian inference on the time of inferior conjunction for planet b is significantly different between the two sectors, with values of 0$\pm$0.008 and 0.81  $\pm$0.49 hours for the light curves of sector 4 and sector 31 with a 2-minute cadence, respectively (the first sector is used as reference). The value and uncertainty derived for sector 4 are similar to those derived by \cite{Winters2019}, who used a different GP kernel. The posterior distribution for the light curve of sector 31 is very broad compared to sector 4 (see Figure \ref{Fig_ttv}). We explain this result by the higher amplitude of the 1.4-day modulation signal (subsection \ref{signal1pt4}) for this sector, which affects the ability of the joint GP-transit model to correctly catch the real value of the time of inferior conjunction T$_0^b$ of planet b. When the inference was made on the combined light-curve dataset, the retrieval of T$_0^b$ was driven by sector 4 light curve, which helps to retrieve the correct joint GP-transit model for the second sector. The uncertainty on T$_0^b$ is only marginally better than the retrieval with sector 4 alone, which indicates that most of the information on the time of inferior conjunction is extracted from the first sector light curve. The posterior distributions of the orbital period of planet b that were retrieved for the two sectors have a similar dispersion, with a small shift of the median value for sector 31 at less than one sigma of the value obtained for sector 4. The orbital period of planet b retrieved with the combined light-curve dataset is well constrained around the median value obtained for sector 4, but with a significant improvement because information is extracted from both sectors. 
A similar effect is visible for planet c. However, the individual transit signals are weaker, which results in a higher uncertainty on the retrieved value of the time of inferior conjunction that dominates any impact of the amplitude difference in the 1.4-day modulation between the two sectors. We obtain the same results when the simple linear detrending approach was used.

\subsubsection{Grazing transit for planet c}
\begin{figure}
\centering
\includegraphics[width=\columnwidth]{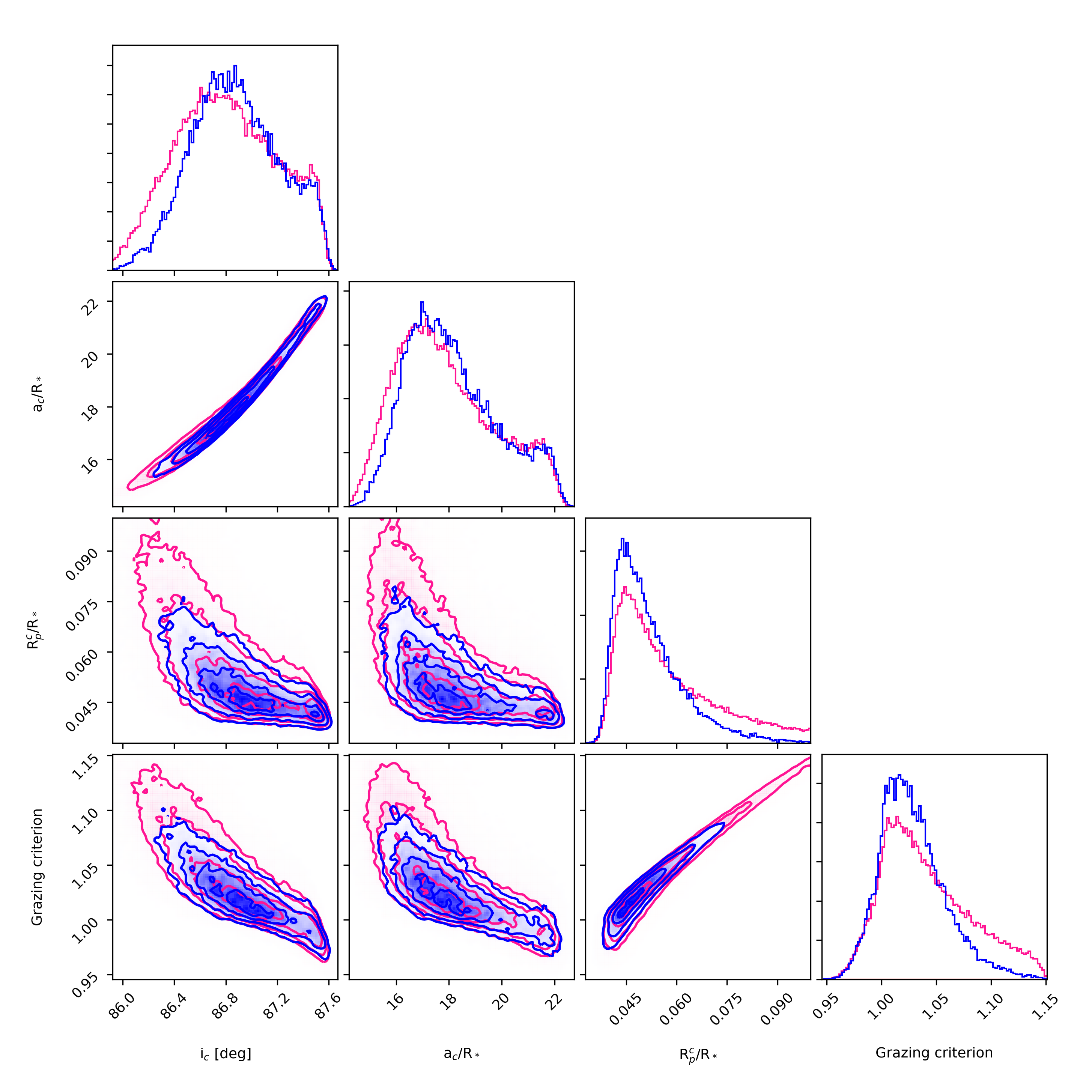}
\caption{\label{grazingplanetc} Corner plot for the photometry analysis (blue) and combined RV-photometry analysis (pink) with the parameters (radius, semi-major axis, and inclination) and the grazing criterion of planet c. A grazing criterion higher than one indicates that the transit is grazing. There is a 85$\%$ chance that the transit is grazing for planet c. }
\end{figure}

The impact parameter of planet c ($b_c = a_c/ R_* * \cos{i_c}$ ) obtained with both approaches (the joint GP-transit model and the linear detrending model) is close to one. Figure \ref{grazingplanetc} shows the posterior distribution of the inclination, the radius, and the semi-major axis retrieved for planet c.
We calculated the grazing criterion R$_p$/ R$_*$ + b, which indicates a grazing transit if it is greater than
one \citep{Smalley_2011}. The distribution (see Figure \ref{grazingplanetc}) peaks around one, with a long tail towards higher values, and we derived a probability of 85$\%$ for the transit to be grazing. The three parameters are strongly correlated. The radius distribution exhibits a long tail towards a high value and up to the boundary of our prior (i.e. 0.1) linked to the possible grazing transit, which is degenerate by nature (i.e. a highly grazing planet with a large radius may have the same transit shape as a smaller planet with a lower grazing transit). 

\subsection{Photometric planetary parameters }\label{GP_TTV}

The period of planet b is refined compared to \cite{Winters2019} to 5.358764$\pm$0.000004 days through the five additional transits we used in this analysis. The retrieved value for the orbital period of planet c is 3.123898$\pm$0.000003.
The complexity of the light curve shown by the sensitivity of the retrieved value for T$_0$ between the two sectors means that the ephemeris and properties derived for the planets are to be considered with caution. Transit-timing variations (TTV) cannot be entirely ruled out. However, the ratio of planet to star radii is marginally affected for planet b. The values derived for the two independent inferences and the combined light curve dataset are compatible within one sigma (Figure \ref{Fig_ttv}). The uncertainty for planet c is higher because the transit is likely grazing, which increases the uncertainty on the radius.

\section{Radial velocity analysis}\label{RVA}
\subsection{Stellar activity}
Radial velocity measurements were derived from the analysis of the host star spectroscopic absorption features. However, the shape and centroid position of these spectral lines are affected by many stellar effects that are commonly grouped under the term 'stellar activity'.  It encompasses a variety of physical processes related to the host star, which affects our capacity of unambiguously analysing the radial velocity signals arising from orbital motion. The activity signals operate on different timescales and amplitudes. They emerge as both incoherent and quasi-periodic variability modulated by the rotation period of the star
\citep{Queloz2011,Cloutier2017}. With a typical stellar jitter magnitude of 1 to 3 m s$^{-1}$ for quiet stars \citep{Dumusque2011} and up to 10 m s$^{-1}$ or higher for more active stars \citep{Santos2014}, it is paramount to investigate the activity effects of LTT 1445A to the best of our ability in order to claim reliable detections of low-mass planets. Unfortunately, these activity signals cannot always be reproduced by analytical models \citep{Fischer2016,Wright2018}. Gaussian processes offer a flexible framework for modelling non-analytical physical processes such as stellar activity and for handling systematic and correlated noise in a robust way. It is broadly used in astrophysics in general \citep{lavie_long_2017,Sale2018,Barros2020} and within the radial velocity community in particular \citep{Haywood2014,Rajpaul2015,ToledoPadr2020,Mortier2020,Faria_2020}.

\subsubsection{Activity indices and Lomb-Scargle periodograms}\label{act_perio}

Spectroscopic observations of LTT 1445A provide several activity indices to probe the stellar activity and rotation of the star. The rotation of LTT 1445A is not seen in the light curve, which indicates either low stellar activity or a rotational period that exceeds the time that is covered by the light-curve observations. The light curve only shows a modulation at 1.4 days, which has been explained by the rotation of one of the binary components of LTT 1445BC. We considered seven activity indices provided by the ESPRESSO pipeline: H$\alpha$, the S-index, contrast, NaD, the full width at half maximum (FWHM), Ca, and R$_{HK}$. As an initial step, the time series of each activity index were cleaned by removing outliers. Activity indicators sometimes correlate with velocities \citep{Fischer2016}.  The radial velocities for LTT 1445A are correlated with the FWHM,  are weakly correlated with the S-index and the NaD index, and are weakly anti-correlated with the contrast index (see Fig. \ref{Ap_TSPcorr}). The Lomb-Scargle periodograms \citep{Lomb_1976,Scargle_1982}  of all indices (see Figure \ref{periodograms_all}) exhibit significant peaks at periods longer than~35 days. The periodograms were computed using the \texttt{astropy} routine \citep{VanderPlas_2012,Astropy_2013,VanderPlas_2015,Astropy_2018}.
The H$\alpha$ and R$_{HK}$ indices also show a significant peak at a short period (between 1.6 and 2 days).
The FWHM index displays the strongest peak in terms of significance at a period of ~72 days.

\subsubsection{Bayesian inference with a GP}\label{BI_GP_RV}

We performed a Bayesian inference with a GP using the same kernel as in section \ref{signal1pt4} independently on all the activity indices (three parameters for each activity index). As noted in the previous section, the choice of kernel was guided by our understanding of the underlying physics that we wished to model. The period hyperparameter controls the rotational period of LTT 1445A, hereafter P$_{rot}$. The length parameter L$_{act}$ is related to the average lifetime of the active regions on the surface of the star. The spot patterns are stable for the duration of the rotational period for M-dwarf stars \citep{Newton_2018b}, therefore this parameter is expected to have a value equal or superior to the period of rotation.
Initially, the inference was made on the logarithm of the hyperparameters with broad uniform priors between -10 and 10 units on all three parameters and without information provided by the periodogram analysis of section \ref{act_perio}. The hyperparameter space was explored with a nested sampling algorithm, as in section \ref{Photo_Tess}. 
The GP converges towards low values for the period parameter and failed to catch the rotational period of the star as revealed with the periodogram analysis. The length parameter also converged to low values for all indices as the GP tried to catch any small variations in the dataset.
In a second approach, we chose a more restrictive prior on the period hyperparameter by forcing it to be higher than 35 days. This reflects the information obtained through the periodogram analysis. We also included an inference on the combined time-series of all activity indices. The inference was made with seven amplitude parameters (one for each activity index) and the two other parameters of the kernel (L$_{act}$ and P$_{rot}$).
Unfortunately, none of these inferences converged to a specific value for the rotational period (Figure \ref{Ap_TSPcorr}). Only the inference with the FWHM index exhibits a weak preference for a period of 72 days.

Finally, we tested models including the C parameter in the GP kernel (from \cite{celerite}). These models are not favoured by the Bayesian analysis. We wish to stress that this result is specific to the LTT 1445A system and should not be considered as a general feature of stellar activity. The analysis of other systems may require the use of the C parameter in the GP kernel.

We conclude that it is not possible to constrain the hyperparameters of the GP with the activity indices for LTT 1445A. Nonetheless, we tried to make an educated guess with the periodogram analysis and our knowledge of stellar activity.

\subsubsection{Expected stellar rotation for LTT 1445A}\label{expectedrot}
The rotation periods of M dwarfs span a broad range of values. \cite{Irwin2011} measured periods from 0.28 to 154 days. Using Kepler observations of M-dwarf stars between 0.3 and 0.55 M$_{\odot}$, \cite{McQuillan2013} reported two distinct groups with periods between 10-25 days and 25-80 days with peaks at $\sim$19 and $\sim$33 days, respectively. \cite{Giacobbe2020} reported rotation periods in the range of 0.52 to 191.82 days with a peak in distribution at $\sim$30 days using the APACHE transit survey of M dwarfs in the mass range 0.15 to 0.70  M$_{\odot}$. The authors also reported a correlation of the rotation period of the star with its mass (when fast rotators with P < 10 days were excluded) and its age.  \cite{Irwin2011} favoured the hypothesis of a rapid spin-down process as a function of age, which seems to be supported by the dearth of stars with intermediate rotation periods of between 10 to 70 days reported by \cite{Newton_2016b,Newton_2016, Newton_2018b}. However, no clear function can help in mapping the rotation period of the star to its age. Estimating the age of stars also remains difficult, which implies systematic high uncertainties on the published values of age.
\cite{Mascareno_2018} proposed a simple equation (equation 1) for M~dwarfs to map the rotational period of the star to the value of its $\log_{10}$R$^{'}_{HK}$. The measured 
$\log_{10}$R$^{'}_{HK}$ for LTT 1445A is -5.88$\pm$0.4, which indicates a rotational period longer than 100 days according this equation. 

It is therefore difficult to estimate the rotation period of LTT 1445A. The periodogram analysis indicates that it is a slow rotator with a period longer than 70 days, which is compatible with the negligible rotational broadening detected by \cite{Winters2019}. We therefore chose to set the rotational period hyperparameter of the GP to the highest peak in the FWHM periodogram at 72 days. The length parameter L$_{act}$ was fixed to 72 days as well in order to take into account that the active regions on the surface of M dwarfs are stable for one or several rotations \citep{Newton_2018b}. We note that the value for P$_{rot}$ is compatible to what we may expect if the system is not too young (see section \ref{stellarprop}). When we assume that the triple star system formed at the same time, a decrease similar to that of LTT 1445A in spin for the other two stars is inconsistent with the 1.4-day signal that results from the rotation of LTT 1445B or LTT 1445C.

\subsection{Identification of radial velocity signals}
In this subsection, we identify potential planets in the radial velocity time-series. The search for radial velocity signals is done without using any information from the light curve analysis performed in section \ref{Photo_Tess}. 

\subsubsection{Radial velocity model for a blind search}
We used only circular orbits, so that one planet was modelled by a Keplerian with only three parameters: the time of inferior conjunction T$_0$, the orbital period P, and the radial velocity semi-amplitude K. The model also includes a mean centre-of-mass velocity for each instrument ($\gamma_{ESP}$ and $\gamma_{HARPS}$) and two parameters to describe the movement of LTT 1445A due to the presence of the BC components: a slope and a curvature.  The Keplerian model can be coupled with a GP with the kernel defined in the previous section. 
The joint GP-Keplerian models were implemented with the RADVEL package \citep{Fulton2018}, which makes use of the celerite package for the GP.

\subsubsection{Iterations with periodogram analysis and Bayesian inference}
\begin{figure}
\centering
\includegraphics[width=\columnwidth]{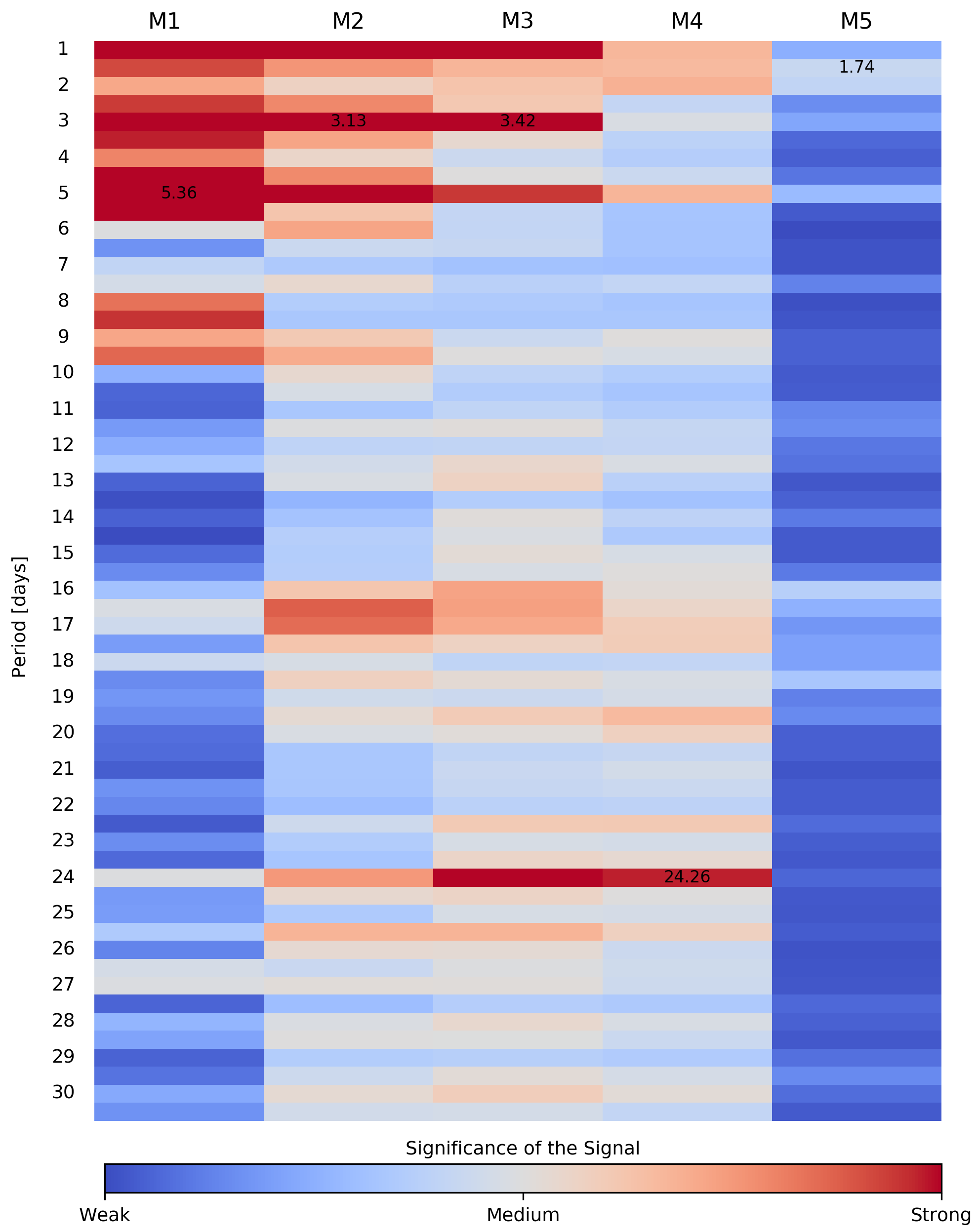}
\caption{\label{hotmap_blindRV} Significance of the Keplerian signals for the blind search in the radial velocity time-series. The x-axis shows the joint GP-RV model we used for the inference from one Keplerian (M1) and up to five Keplerians (M5). The y-axis indicates the subdivision of the orbital period parameter space with a bin size of 0.5 days cut at 30 days. One cell corresponds to a pair model or bin that indicates the model we used for the inference and the prior on the orbital period of the planet. The logarithm of the ratio of the Bayesian evidence (Bayes factor) between the pair model (M$_x$)/bin and the best pair model (M$_{x-1}$)/bin is colour-coded from weak (value of 0) to strong (value of 15 or above) to indicate the significance of the signal detected in the corresponding bin (except for the first column, in which the Bayes factor of each pair M1/bin is computed with the worst pair M1/worst bin). The orbital period obtained for the best signal of each iteration is written in black in the corresponding cell. An example: the cell corresponding to the M2 joint GP-two Keplerian model for the period subdivision of 3 to 3.5 days (the retrieved period is 3.13 days) indicates a strong significance of this signal compared to a GP-RV model with only one Keplerian at 5.36 days, which was the best signal detected with the previous iteration with M1.}
\end{figure}

We explored the different signals in the radial velocity time-series by subdividing the period parameter space into 200 bins of 0.5 days. We performed a Bayesian inference within each bin with the simple circular Keplerian (from one planet without GP to five planets with GP). One iteration corresponded to the inference made on all the period bins with one model. The prior on the orbital period was set to be uniform within each bin. The priors on the other parameters are given in Table \ref{SumTable}. The multi-dimensional parameter space was explored with the nested sampling algorithm, which provided the Bayesian evidence that was used to classify the different periodic signals in the radial velocity time-series. In the first iteration, the strongest signals found with the one-planet model were at higher periods close to the possible rotation of the star at 72 days. For comparison with this period-subspace model ladder, we also computed the LombScargle periodogram using the ASTROPY implementation in Python \citep{Astropy_2013,Astropy_2018} for the ESPRESSO dataset alone, which confirmed that the main signal is compatible with the 72-day signal seen in the FWHM activity index. The second iteration was made with a joint GP-one Keplerian model. The hyperparameters L$_{act}$ and P$_{rot}$ of the GP were fixed to the value chosen in section \ref{BI_GP_RV} (i.e. 72 days), only the amplitude was allowed to vary. In the next iteration, we fixed the parameters from the most significant signal in the previous iteration, added another planet, and proceeded with another inference on each bin. We iterated this process until no clear signal remained in the residue. 

The subdivision of the orbital period parameter space and the strategy to fix the values of each signal before adding new Keplerian signals allowed a quick scan of the parameter space. In each iteration, the bulk of the posterior was found more easily with a reduced parameter space and fewer free parameters. The computed Bayesian evidence for each model (M1, M2, etc.) is a good proxy for the true value of these models with the larger period parameter space because the different bulks of the posterior of each significant signal are the main contribution to the evidence integral.

\subsubsection{Identification of the different signals}

The result of this procedure is summarised in Figures \ref{hotmap_blindRV} and \ref{blindRV_App}. The periodogram analysis yields results that are similar to those of the Bayesian inference for the first four iterations, but then failed to identify the 24-day signal. Figure \ref{hotmap_blindRV} shows the significance of the signal detection within each orbital period bin for all the joint GP-Keplerian models with one to five planets. The significance is evaluated as the logarithm of the ratio of the Bayesian evidence (Bayes factor)  of each pair inference (model/bin) with the best pair inference of the previous iteration. The Bayes factor for each bin with the joint GP-one Keplerian model (column M1)  was computed with the lowest evidence value amongst the different period subdivision with the same model because any bin is strongly supported compared to the highest evidence obtained with the one-Keplerian model without GP.

\paragraph{Confirmation of the two transiting planets.}
The first two detected signals are at 5.36 and 3.13 days, which corresponds to the orbital periods obtained for planet b and c with the photometry analysis in section \ref{Photo_Tess}. This is a direct confirmation that the two planets orbit the same host star LTT 1445A. 

\paragraph{A complicated signal at 3.4 days.}
The third significant Keplerian signal is at a puzzling period of 3.4 days. A planet with an orbital period like this will have repercussions on the dynamical properties of the system and especially on the orbit of planet c with a potential TTV, which we do not see. The analysis of the light curve did not show any potential transits with a 3.4-day period, which implies a planet inclined compared to the 5.36- and 3.13-day planets. It therefore appears to be difficult to claim that this signal is due to the presence of another planet. We note that a 3.4-day period is an alias of the 1.4 days in our time-series, which is the value of the modulation signal seen in the light curve that we have associated with either B or C. Therefore, we suspected that this signal was a contamination of some light from LTT 1445BC that entered ESPRESSO. However, the separation between LTT 1445A and the BC component is about 7" \citep{Mason_2001,Dieterich_2012} compared to the one-arcsecond diameter of the fiber of ESPRESSO \citep{Pepe_2014,Pepe2021}. The average seeing of Paranal is 0.77", which will place the LTT 1445BC light source at several FWHM from the fiber. The contamination is therefore strongly unlikely. Nonetheless, we tested the hypothesis of a contamination from the 1.4-day modulation by performing the same blind search with the two GP kernel hyperparameters P$_{rot}$ and L$_{act}$ fixed to the value obtained with the photometry analysis. The long-period signals bound to the rotation period of the star are partially caught by this short-period GP because GPs with exponential decay components in their kernels destroy the covariance between data points that are separated by much more than the decay timescale, so they can model signals at lower frequencies than their exponential-decay timescales. The signal with the highest significance remains the orbital period of 5.36 days of planet b (see Figure \ref{hotmap_blindRV_1pt4}). In the second iteration (M2 model), the signal of planet c at 3.13 days is detected only weakly. We added a Keplerian for planet c with the ephemeris obtained with the photometry analysis. The third signal is the 24-day signal presented below, which is also seen in the blind search with the 72-day GP kernel (it is the most significant signal for the inference with the M2 model). The other iterations with the more complex models M4 and M5 only show weak Keplerian signals. The 3.4 days disappeared completely. To test our hypothesis further, we fit a joint GP-Keplerian model at 1.4 days with the GP at 72 days, which also removed the 3.4-day signal without weakening the significance of the two transiting planets and the 24-day signal presented below. The origin of the 3.4-day signal remains unknown. A full dynamical analysis of the system is beyond the scope of this paper, but we stress that this analysis is necessary to reliably rule out the possibility of another small planet with an orbital period of about 3.4 days. We note that which hypothesis is chosen to include the signal in the models does not impact the parameters that are derived for the three other planets.
  
\paragraph{A possible planet at 24 days}
The fourth significant Keplerian signal is at a 24-day orbital period with a semi-amplitude of $\sim$1.48 m.s$^{-1}$ . It does not correspond to any peak in the periodogram analysis of the stellar activity indices (Figure \ref{Ap_TSPcorr}). We searched for potential transits using the ephemeris provided by the inference on the radial velocity dataset, but did not find any signal. However, only one transit for each sector would be visible, with the risk that the transit for sector 4 falls within a gap in the light curve based on the radial velocity ephemeris. This low number of potential transits combined with the 1.4-day modulation signal complicates any detection of the planet in the light curves, so it is not possible to entirely rule out the possibility that a potential planet (with the ephemeris retrieved from the RV) does transit.
\medbreak
\medbreak


\subsection{Radial velocity analysis}
In this subsection we perform a thorough analysis of the radial velocity time-series with different models and use the identification of the different Keplerian signals from the quick blind search. In this subsection, information from the light curves analysis are used as prior for the two inner planets detected with both observing methods. 

\subsubsection{Models and nomenclature}

We considered a set of models that encapsulates different assumptions to model the Keplerian signals in the radial velocities. 

Firstly, one model included a certain number of planets, indicated by the first number after the letter M (M1, M2 etc.). Each planet is modelled by a Keplerian with three parameters for circular orbits (the time of inferior conjunction T$_0$, the orbital period P, and the radial velocity semi-amplitude K) or with five parameters for eccentric orbits (eccentricty e and longitude of periastron $\omega$ are added to the free parameters). The assumption for each planet is denoted by the letters following the notation of the model, a letter c or e per planet indicates circular or eccentric orbits, respectively. 

Secondly, each model included a mean centre-of-mass velocity for each instrument ($\gamma_{ESP}$ and $\gamma_{HARPS}$) and possibly a jitter term for each instrument ($\sigma_{ESP}$,$\sigma_{HARPS}$). These assumptions are denoted by the letter G when no jitter is added to the model and by a J when both the $\gamma$s and $\sigma$s are included in the model. The additional jitter terms aim at modelling any source of white noise in the radial velocities that will not be caught by the joint GP-model or the default uncertainties from the pipeline. The jitter terms are added in quadrature to the formal uncertainties derived by the ESPRESSO and HARPS pipelines.

Thirdly, three assumptions were considered for the motion of LTT 1445A due to the presence of the BC components: a slope, a slope and a curvature, or no movement (no additional parameters). The slope is denoted in the model name with the letter L  and the curvature by the letter C. When the name contains no letter, then the model does not include any parameters for the movement of LTT 1445A. It is very unlikely that models without a slope are favoured by the model selection because the trend is clearly visible by eye in the radial velocity time-series. We tested it regardlessly to evaluate the significance of the two other assumptions.

Fourthly, the use of a joint GP-RV model is indicated by the notation gp1 in the name of the model. We  used the kernel defined in section \ref{identification_transit} with the two parameters P$_{rot}$ and L$_{rot}$ fixed to 72 days, as discussed in section \ref{expectedrot} . Only the amplitude of the GP is a free parameter.

Finally, we chose to model the 3.4-day signal in the radial velocities with either a Keplerian with an orbital period of 3.4 days (the third signal in models with at least three Keplerian) or 1.4 days (the third signal in models with at least three Keplerian and a superscript $^{1.4}$ in the name of the model), or with a GP with its hyperparameters fixed to the photometry values obtained for the 1.4-day modulation signal in the light curve. Only the amplitude of the GP is a free parameter. When the model uses this assumption, the model name includes gp2  and does not possess the 72-day GP.

\subsubsection{Bayesian inference and model selection}
\begin{figure}
\centering
\includegraphics[width= \columnwidth]{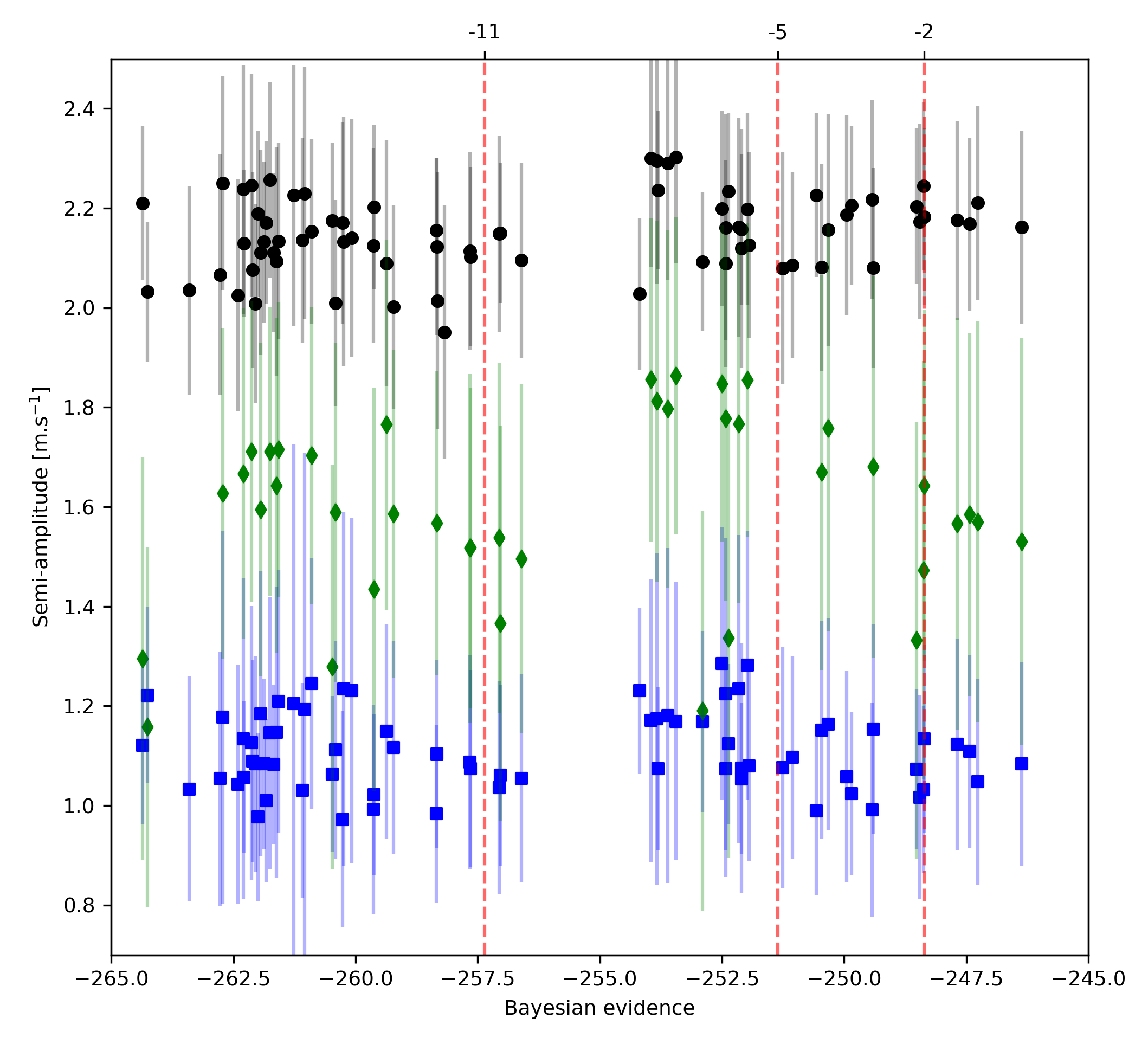}
\caption{\label{k_vs_logz} Radial velocity semi-amplitude of planets b (black), c (blue), and d (green) as a function of the logarithm of the  Bayesian evidence obtained with the inference on the radial velocity dataset alone. The higher the Bayesian evidence, the more favoured the model. The vertical dashed red line indicates the different thresholds in the difference with the logarithm of the Bayesian evidence of the best model. Models with a difference larger than 5 are strongly rejected.}
\end{figure}

\begin{table*}
\begin{center}
\caption{\label{modelladderreduced} Model ladder for the radial velocity analysis.}
\begin{tabular}{ccccccc}
\hline
Model name &   K$_b$  & K$_c$  &  K$_d$  & ln $\mathcal{Z}$ & Bayes Factor$^{1}$  & Bayes Factor$^{2}$ \\
& m.s$^{-1}$ & m.s$^{-1}$ &  m.s$^{-1}$ & & (log) & (log) \\
\hline
\hline
\multicolumn{7}{c}{Models with the 72 days GP}\\
 \hline
M4ccecJLgp1 &   2.16  & 1.08  & 1.53 & -246.36  & 0.90 & Best model\\
M4eeeeJLgp1 &   2.21  & 1.05  & 1.57 & -247.26  & 0.17 & -0.90\\
M5cceccJLgp1 &   2.17  & 1.11  & 1.58 & -247.43  & 0.25 & -1.07\\
M4ccccJLgp1 &   2.18  & 1.12  & 1.57 & -247.68  & 0.68 & -1.32\\
M5cccccJLgp1 &   2.18  & 1.13  & 1.64 & -248.36  & 0.02 & -2.00\\
M4eeeeGLgp1 &   2.24  & 1.03  & 1.47 & -248.38  & 0.08 & -2.02\\
M3cceJLgp1 &   2.17  & 1.02  & - & -248.45  & 0.06 & -2.09\\
M4ccecGLgp1 &   2.20  & 1.07  & 1.33 & -248.51  & 0.89 & -2.15\\
M4ccecJLgp1$^{1.4}$ &   2.08  & 1.15  & 1.68 & -249.40  & 0.03 & -3.04\\
M3eeeJLgp1 &   2.22  & 0.99  & - & -249.43  & 0.42 & -3.07\\

\hline
\multicolumn{7}{c}{Models with the 1.4 days GP}\\
\hline
M4ccccGLgp2 &   2.20  & 1.28  & 1.86 & -251.98  & 0.18 & 0.18\\
M3cccGLgp2 &   2.16  & 1.23  & 1.77 & -252.16  & 0.27 & Best model\\
M3cccJLgp2 &   2.16  & 1.22  & 1.78 & -252.42  & 0.08 & -0.27\\
M4ccccJLgp2 &   2.20  & 1.29  & 1.85 & -252.50  & 0.95 & -0.34\\
M4eeeeGLgp2 &   2.30  & 1.17  & 1.86 & -253.45  & 0.16 & -1.29\\
M3eeeGLgp2 &   2.29  & 1.18  & 1.80 & -253.61  & 0.23 & -1.45\\
M3eeeJLgp2 &   2.29  & 1.17  & 1.81 & -253.83  & 0.12 & -1.67\\
M4eeeeJLgp2 &   2.30  & 1.17  & 1.86 & -253.96  & 6.13 & -1.80\\
M2ccGLgp2 &   2.14  & 1.23  & - & -260.08  & 0.16 & -7.92\\
M2ccJLgp2 &   2.13  & 1.23  & - & -260.24  & 0.66 & -8.08\\
\hline
\hline
\multicolumn{2}{l}{$^{1}$ ln $\mathcal{Z}$$_i$ - ln $\mathcal{Z}$$_{i-1}$}\\
\multicolumn{2}{l}{$^{2}$ ln $\mathcal{Z}$$_{i-1}$ - ln $\mathcal{Z}$$_{best}$}\\

\end{tabular}
\tablefoot{The results are showed between the two approaches for the 3.4 days signal (alias of 1.4 days): a Keplerian (period of the alias at 3.4 days or period at 1.4 days, which is models with $^{1.4}$) or a GP with the hyperparameters fixed at the retrieved values of the photometric analysis. The logarithm of the Bayes factor is reported for each pair of consecutive models and for each model with the best model (M4ccecJLgp1 for the 72 days GP and M3cccGLgp2 for the 1.4 days GP).
}
\end{center}
\end{table*}

We performed the Bayesian inference with each model on the radial velocity time-series using the nested sampling algorithm with the same setup as in the previous sections (1000 live points, 0.8 sampling efficiency, and at least three multiple runs).
The first two planets in all the multi-planet models was planet b, and planet c was also detected in the photometry analysis. The priors on the time of inferior conjunction and the orbital periods of these planets were Gaussian, with the centre and the standard deviation set to the retrieved values of the inference on the light curve (section \ref{Transit_res}). The constraints on the ephemeris obtained with the photometry analysis are superiors to the one obtained with the radial velocities alone. 
For the other signals, the prior on the orbital period and the time of inferior conjunction were uniform in a range that encompassed the period obtained with the blind search. The priors were all uniform on the other parameters and were the same as in the combined analysis (section \ref{Discu}) given in Table \ref{SumTable}. 

The first output of our analysis is the model ladder. It classifies the significance of all models using Bayesian evidence. The full table with all the models is shown in Table \ref{tadder}, and a reduced table is presented in Table \ref{modelladderreduced}, in which the best ten models are broken down between the two hypothesies on the 3.4-day radial velocity signal (a GP or a Keplerian). Our threshold for weak, moderate, strong, and high significance of a model are with logarithms of the Bayes factor of 2, 3, 5, and 11, respectively \citep{Trotta:2008aa}. A model with more complexity (i.e. more parameters) needs at least an increase in logarithm of the evidence of two compared to a more simple model (i.e. with fewer parameters) to be moderately significant and favoured.
From this analysis, we can derive the following conclusion for the LTT 1445A bcd planetary system: 

\paragraph{GP models are favoured.} Models with a GP(with period of 1.4 or 72 days) are strongly favoured because they absorb the main significant signals in the radial velocities at long periods that we assume to correspond to the stellar rotation. We point out that our analysis is skewed towards planets with small periodicity (i.e. less than ~ 40 days) because of the time span of our dataset.  

\paragraph{Planets b and c strongly favoured.} The existence of the two planets b and c is strongly favoured by the model selection. Models with eccentric orbits (MXee* with X $\geqslant$ 2) have no or a weak significance compared to the respective model with circular orbits for both planets (MXcc*), which indicates that the two additional parameters of the eccentric models add complexity that is not required to explain the data. More observations (or more accurate data) should help to constrain the eccentricities better.

\paragraph{3.4 day signal.} It is necessary to model the 3.4-day signal either by a Keplerian orbit (models M4 with gp1) or by the 1.4-day GP (models M3 or M4 with gp2). Models with a Keplerian at 3.4 or 1.4 days have a higher evidence than models with the 1.4 days GP because the long-term variations are better handled by the 72-day GP.

\paragraph{Long-term variation.}  The model selection favours models with jitter terms (letter J) and a slope (letter L) in the case of the 72-day GP. Jitter terms are not favoured in the case of the 1.4-day GP because this latter absorbs any short-period jitter more efficiently than the 72-day GP. The curvature parameter is not favoured. This could indicate that LTT 1445A is in the linear regime of its orbit with the components BC or that the curvature is too small and is entirely dominated by the short-term variations (stellar jitter and planets).

\paragraph{A new planet candidate.}  The 24-day signal is moderately favoured by the model selection when the 3.4-day signal is modelled with a Keplerian (the logarithm of the Bayes factor between M4ccecJLgp1 and M3cceJLgp1 is 2.09) and strongly favoured in the case of the 1.4-day GP models (ln $\mathcal{Z}$$_{M3cccGLgp2}$ - ln $\mathcal{Z}$$_{M2ccGLgp2}$ = 7.92).  The detection under the other assumptions (slope, curvature, and jitter) is in approximately the same range of significance. Therefore, we claim the existence of planet d orbiting LTT1445 A with a period of 24 days with a moderate significance. We advocate for more observations to bolster the significance of the detection.

The retrieved values for the semi-amplitude of the three planets are robust as they are stable to the different assumptions, except for models that failed to explain the data (with Bayesian evidence smaller than minus 300). Figure \ref{k_vs_logz} shows the different values obtained for the best models with the different thresholds of the model selection. 
Finally, we ran retrievals with the four Keplerian signals, the slope and jitter (model M4ddcdJL), but with a Gaussian prior on the two other hyperparameters P$_{rot} $and L$_{act}$ of the GP. We tested two Gaussian priors: 72 $\pm$10 days on both P$_{rot} $ and L$_{act}$ based on the FWHM value, and 130 $\pm$30 days based on the derived value from \cite{Mascareno_2018}. We also performed a retrieval with a GP including the C parameter as a free parameter and the same setup as above. The retrieved distributions for the Keplerian parameters are not affected by these new hypotheses.



%

\begin{figure}
\centering
\includegraphics[width= 0.75\columnwidth]{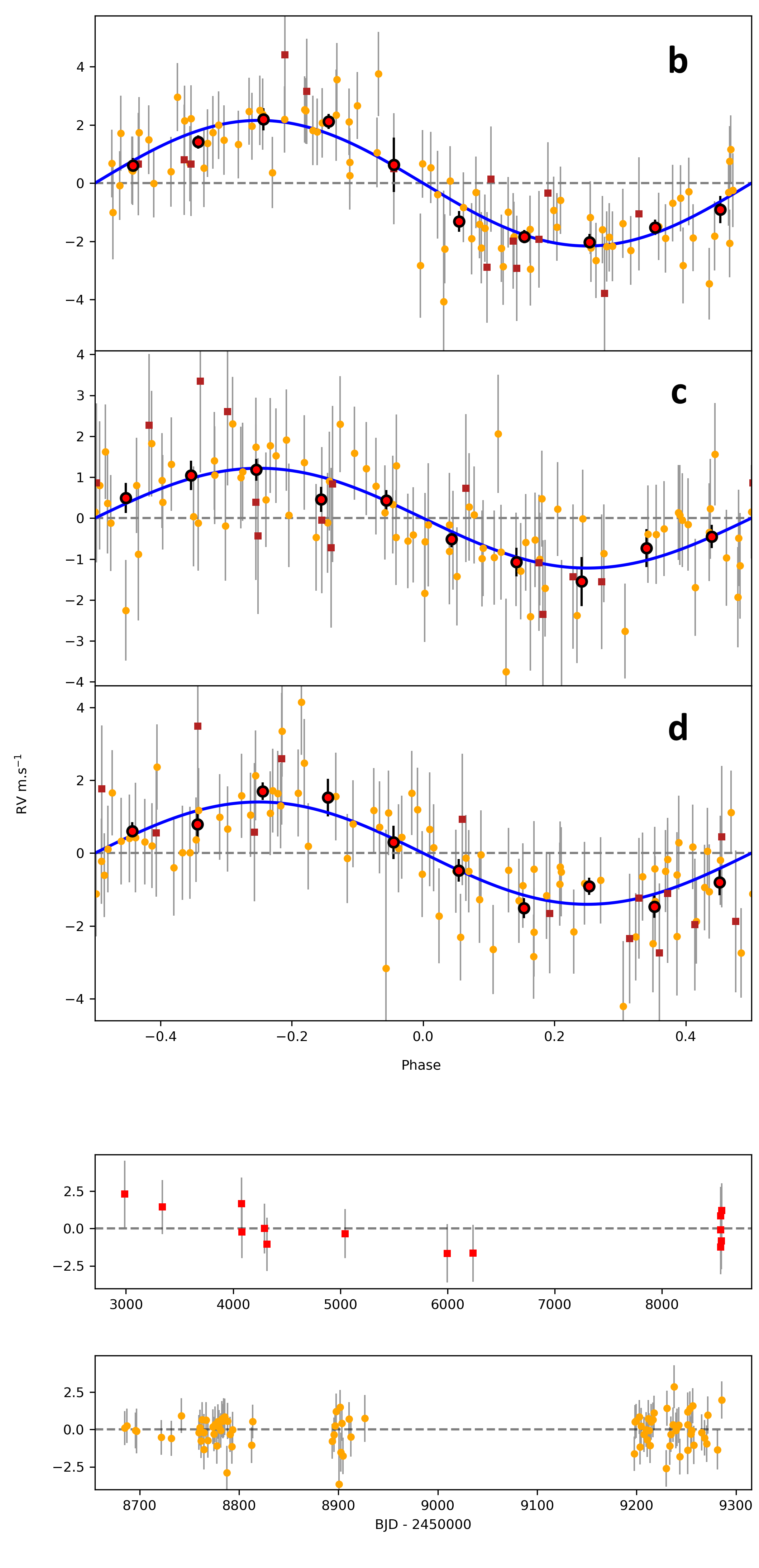}
\caption{\label{RVbestfit} Phase-folded radial velocities for planets LTT 1445A b (top panel), c (second panel), and d (third panel). The red squares are HARPS data, the orange circles are ESPRESSO data, and the red circles are the ESPRESSO data binned in ten bins. The blue lines indicate the best fit obtained with the combined photometric and radial velocity datasets. The two last panels show the residuals for HARPS observations (second to bottom panel) and ESPRESSO (bottom panel).  }
\end{figure}

\begin{figure}
\centering
\includegraphics[width=\columnwidth]{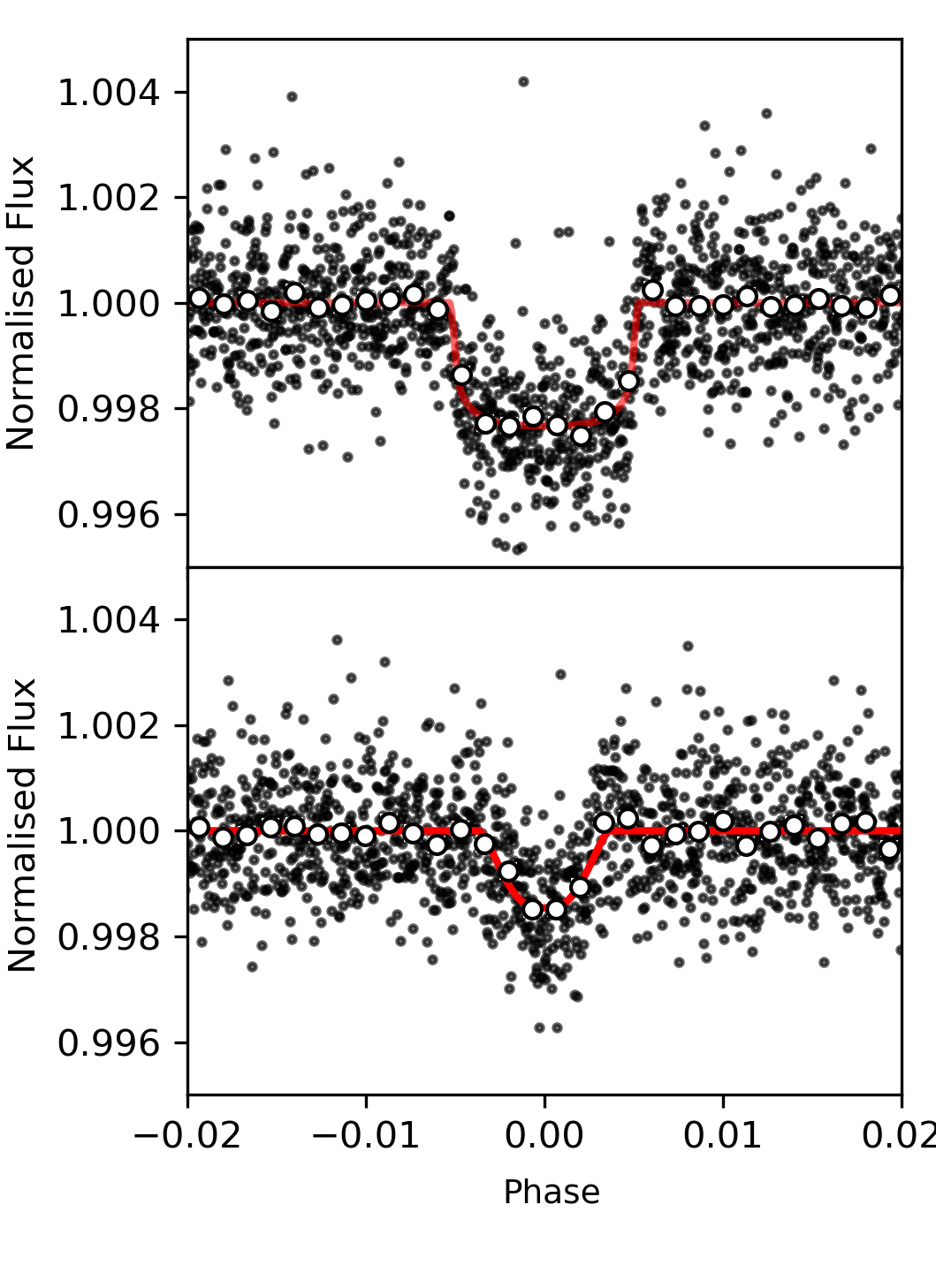}
\caption{\label{lc_bestfit} Phase-folded light curves with both sectors at a 2-minute cadence for planet b (top panel) and planet c (bottom panel). The black dots show the flux for the 2-minute cadence detrended with the GP from the best-fit model obtained with the combined photometric and radial velocity datasets. The white dots with small error bars are the same flux binned. The red line shows the transit model component of the best fit.}
\end{figure}
 
\section{Combined RV-photometry analysis}\label{Discu}
As a final step of our analysis, we performed a joint inference on the photometric and spectroscopic datasets to derive self-consistent uncertainties taking the correlations between the different parameters of the light curve and radial velocity models into account (similar to \citealt{Sozzetti_2021}). We used the best radial velocity model for the circular (M4ccecJLgp1) and eccentric (M4eeeeJLgp1) orbits with the respective photometric model with both transiting planets. We only used the 2-minute cadence for both sectors for the photometric dataset. The results are consistent with the individual radial velocity and TESS analysis. Table \ref{SumTable} shows the values for LTT1445A planetary system retrieved with this joint inference. 
As a summary, we confirm the discovery of planet LTT 1445A b \citep{Winters2019}  with a refined orbital period of 5.358764$\pm$0.000004\ days around component A of the system of three M dwarfs LTT 1445ABC at 6.9 pc. We present the discovery of LTT 1445A c with a period of 3.123898$\pm$0.000003 days and the candidate planet LTT 1445A d at 24.30$^{+0.03}_{-0.08}$ days. The two inner planets are transiting, but there is a high probability that the transit of planet c is grazing. We derived radii of 1.43$\pm0.09$$\Rearth$ and 1.60$^{+0.67}_{-0.34}$$\Rearth$ for planets b and c, respectively. 
We found no evidence that LTT 1445A d transits, but the time window covered by the two TESS sectors and the strong impact of the 1.4-day modulation signal on the light curves prevent us from ruling this possibility out. The masses derived from this analysis are 2.32$\pm$0.25$\Mearth$, 1.00$\pm$0.19 $\Mearth$ for planets LTT 1445A b and c, respectively. We derived a minimum mass for the potential planet d of 2.72$\pm$0.75  $\Mearth$.
Furthermore, LTT1445A exhibits a long-term drift. We derived a radial velocity slope of -4.4 m.s$^{-1}$.yr$^{-1}$, which agrees with the maximum expected value if it is due to the gravitational of BC, which according to \cite{Torres_1999}, given the total mass of BC provided by \cite{Winters2019}, the system distance and present-day separation of around 7", is 9 m.s$^{-1}$.yr$^{-1}$. Furthermore, \cite{Kervella_2019} reported a statistically significant (almost 20 sigma) proper motion anomaly at the mean epoch of Gaia DR2 for LTT 1445 A. Finally, Gaia EDR3 lists low RUWE values (1.07)  and astrometric excess noise (158 microarcsec). All the points above confirm that the motion of the primary is due to the outer BC components, and a closer-in (within about 5 au) massive gas giant or brown dwarf companion can be ruled out.

\begin{table}
\begin{center}
\caption{\label{PhysicTable}Physical properties of the planetary system LTT 1445A }
\begin{tabular}{lcc}
\hline
Parameters  & Units &  Values  \\
\hline
\multicolumn{2}{c}{LTT 1445A} & \\
M$_*$ & Mass (M$_\odot$) & 0.249$\pm$0.023\\
R$_*$ & Radius (R$_\odot$) & 0.276$ \pm$0.010\\
& & \\
\multicolumn{2}{c}{planet b} & \\
P$_b$ & Period (days) & 5.358764$\pm$0.000004\\
R$_b$ & Radius ($\Rearth$) & 1.43$\pm$0.09 \\
M$_b$ & Mass ($\Mearth$) &2.32$\pm$0.25 \\
a$_b$ & Semimajor axis (AU) & 0.032$^{+0.006}_{-0.003}$\\
$\rho_b$ & density (g.cm$^{-3}$) & 4.36$\pm$0.74 \\
& & \\
\multicolumn{2}{c}{planet c}   & \\
P$_c$ & Period (days) & 3.123898$\pm$0.000003 \\
R$_c$ & Radius ($\Rearth$) & 1.60$^{+0.67}_{-0.34}$ \\
M$_c$ & Mass ($\Mearth$) &1.00$\pm0.19$ \\
a$_c$ & Semimajor axis (AU) &  0.0226$\pm$0.0040\\
& & \\
\multicolumn{2}{c}{planet d}  & \\
P$_d$ & Period (days) & 24.30$^{+0.03}_{-0.08}$ \\
Msini$_d$ & Mass ($\Mearth$) &2.72$\pm$ 0.75\\
a$_d$ & Semimajor axis (AU) & 0.09$\pm$0.02\\

\end{tabular}
\end{center}
\end{table}

\section{Discussion}
\begin{figure}
\centering
\includegraphics[width= \columnwidth]{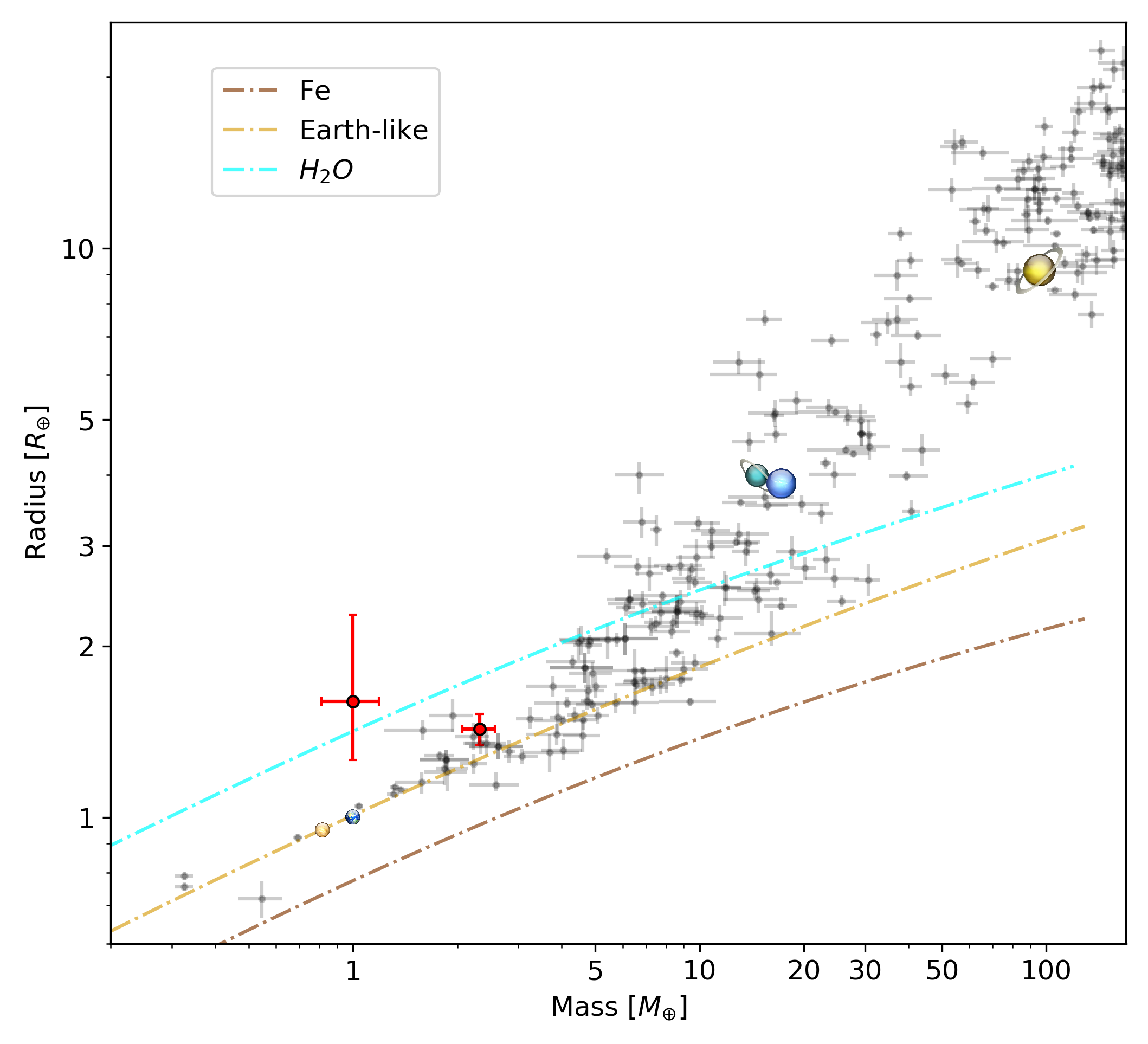}
\caption{\label{MRdiagram} Mass-radius diagram including the planets selected by \cite{Otegi_2020}. The two inner planets LTT 1445Ab and LTT 1445Ac are indicated in red.  }
\end{figure}

Figure \ref{MRdiagram} shows that planet b is compatible within two sigma with an Earth-like composition \citep{Dorn_2015}. The planet is less dense than Earth, with a density of 4.36$\pm$0.74 g.cm$^{-3}$, which might indicate the presence of an atmosphere. Alternatively, the planet may have lower iron-mass fractions because the host star is metal poor \citep{Adibekyan_2021,Santos_2017}. Follow-up observations focusing on the detection and characterisation of the atmosphere of the planet are needed to explore these different hypotheses. The uncertainties on the radius of planet c are high because the transit may be grazing. The planet may be compatible with a water-rich composition, but any real statement on the nature of the planet seems premature. The planet is also compatible at two sigma with an Earth-like composition. Future radial velocity observations combined with photometric observations for example with CHEOPS \citep{Benz_2021} should reduce the uncertainties to allow a clearer picture of the density of this planet. The amount of water is expected to be low for planet b in comparison to the refractory materials (e.g. iron or silicates). Because the two planets are on compact orbits, if planet c were water-rich, it would require a mechanism to bring water to it without enhancing b or a mechanism to deplete water on planet b without depleting the water on planet c. Alternatively, planet c may have formed beyond its current orbit far away in the disc and then migrated to its current position. In that case, its water content may be very different from that of planet b. This seems unlikely, however, as any migration of planet c would have consequences on planet b as well.
In any case, the presence or absence of an atmosphere cannot be unambiguously claimed for either planet.  As discussed by \cite{Winters2019}, it is unlikely that we will find better targets than LTT 1445A bc based on the known occurrence rates of planets around M dwarfs. The system is therefore a prime target for an atmospheric characterisation with space telescopes such as the James Webb Space Telescope (JWST).
 \cite{Winters2019} used \cite{Kopparapu_2013,Kopparapu_2014} to compute the conservative inner and outer habitable zone (HZ) boundaries for LTT 1445A and found 0.093 and 0.182 AU, respectively, which places planets b and c far away from these limits. With a semi-major axis of 0.09 AU, the super-Earth LTT 1445Ad may sit within the inner boundary of the habitable zone. The habitability around M dwarfs has been the subject of extensive debate, in particular, with regard to the prolonged stellar activity or the gravitational effects that result in tidally locked configuration for planets within the habitable zone (see \citealt{Shields_2016} and references within for an overview). The habitability of planets in multiple stellar system also brings its share of difficulties, especially in terms of orbit stability. Many studies have been devoted to the habitability in binary systems \citep{Jones_2001,Noble_2002,Takeda_2008,Dvorak_2010}. In principle, these studies can also be expanded to systems of higher order \citep{Cuntz_2014}. The interest in the habitability of LTT 1445A d is limited with the transit technique, if indeed it does not transit. However, the planet may be a prime target for next-generation instruments such as the ELT-HIRES with its adaptive optics that may allow us to characterise a possible atmosphere. 

Of the more than 3'500 planetary systems known thus far, 111 are planetary systems around binary stars, and 26 lie in multiple star systems \citep{Schwarz_2016}. 
Unfortunately, we only know a handful of multi-planetary systems around triple or higher-order star systems so far. A radial velocity detection of up to seven planets has been claimed for the M dwarf GJ 667 C \citep{Anglada_2012,Gregory_2012}. The star is a companion of the K3V+K5V binary GJ 667AB at a minimum separation of ~230 au. Two Jovian planets with eccentric orbits have been detected by radial velocity around the main component of two triple systems, HD 207832 A \citep{Haghighipour_2012,Ngo_2017} and HD 65216 A \citep{Mayor_2004,Mayor_2011}. Both hosts are G-type stars and possess two companions at projected separations of 110 and 253 AU, respectively.
Finally, two planetary systems with radial velocities and transit detection have been reported for the F star K2-290 \citep{Hjorth_2019,Hjorth_2021} and the K star Kepler-444 \citep{Campante_2015}. The first system possesses a close-by companion at 113 AU and a far-away companion at more than 2450 AU. The two planets are a hot mini-Neptune and a warm Jupiter.  Kepler-444 possesses two unresolved M-dwarf companions at a separation of 66 AU \citep{LilloBox_2014,Dupuy_2016} with very eccentric orbits that bring these two companions very close to the main component (within 5 AU). Kepler-444 is also known to be the oldest planetary system with five compact (within 0.1 au) small planets. In this context, the report of the LTT 1445 A bcd planetary system is a good addition to this small sample, but is not sufficient to derive robust statistics regarding multi-planet systems around triple or higher-order star systems. We note that LTT 1445 is the only known system composed of three M dwarfs. The architecture is similar to that of Kepler-444, HD 207832, and HD 65216 in the sense that the planets have been detected around the main component of the star system and that the host possesses two companions that orbit each other closely (6 AU for HD 65216 BC, 4 AU for HD 207832 BC, and unresolved for Kepler-444 BC). Current literature on exoplanets in multiple star systems focuses on binaries because on one hand, they are the main systems among the multiple stellar systems, and on the other hand, many more planets have been detected around them than around higher-order systems.
It appears from observations that the dynamical properties, in terms of eccentricity, mass, and period distribution of planets orbiting binaries are not different than those of planets around single stars \citep{Ngo_2017} in the case of giant planets between 0.1 and 5 AU, but extrapolation to smaller planets is not trivial.
The presence of a companion or a binary companion as in the case of LTT 1445 A has strong implications for planet formation mechanisms as the protoplanetary disc may be truncated \citep{Artymowicz_1994} and conditions will be less favourable for planet formation \citep{Nelson_2000,Mayer_2005}. Stellar companions may also eject any formed planets \citep{Zuckerman_2014}. The three planets LTT 1445A bcd are S-type planets because they orbit one component of the stellar multiple system \citep{Dvorak_1986}. The system is sufficiently compact (within 0.2 AU) to be dynamically stable when the work done on binaries is extrapolated  \citep{Holman_1999,Marzari_2016,Cesare_2021}  to LTT 1445A -BC and a separation of 21 AU between the two components is assumed \citep{Winters2019}. The stability of planet orbits around binaries is strongly dependent on the eccentricity of the companion. The orbit of LTT 1445 BC is unknown, and we cannot rule out the possibility that a high eccentricity will bring the two companions closer to the host star. However, stars may retain their planets if the system is compact enough even if companions are on very eccentric orbits. For example, Kepler-444 possesses a binary companion on a very eccentric orbit that regularly brings it to $\sim$5\,AU of the primary star. Nevertheless, five planets were found to orbit the primary, and recent study showed the poor dynamical perturbative impact of the stellar binary on the stability of the planetary orbits \citep{Stalport_2022}. The entire LTT 1445 system may be coplanar \citep{Winters2019}, and the stability of the planetary system may also reflect some particular aspect of the formation history of LTT 1445ABC.   

To conclude, the LTT 1445 system offers great opportunities for studying a multi-planet system in a triple star system with two terrestrial planets that transit and a super-Earth in the habitable zone. We advocate for spectroscopic and photometric follow-up observations of the system to confirm the status of planet d, to prepare an atmospheric characterisation of the two inner planets with the next generation of telescopes, and to monitor the presence or absence of other planets in the system.

\setlength{\extrarowheight}{0.15cm}
\begin{table*}
\tiny
\begin{center}
\caption{\label{SumTable} Prior and posterior distributions for the combined photometric and spectroscopic analysis for the LTT 1445A planetary system. }
\begin{tabular}{lccc}
\hline
Parameter (unit)  & Prior & Posterior (circular) & Posterior (eccentric)     \\
\hline
\multicolumn{4}{c}{planet b}\\
Inferior conjunction (T$_0$) [BJD - 2457000]   &  $\mathcal{U}$(1421,1426)  & 1423.426464$^{+0.000351}_{-0.000350}$ &  1423.426378$^{+0.000299}_{-0.000304}$\\
Orbital period (P) [days]   &  $\mathcal{U}$(5,6) & 5.358764$^{+0.000004}_{-0.000004}$  &  5.358765$^{+0.000003}_{-0.000003}$\\
Radius in stellar radii (R$_p$/R$_*$)   &  $\mathcal{U}$(0.02,0.1) &  0.04740$^{+0.00126}_{-0.00138}$  &  0.04618$^{+0.00085}_{-0.00086}$\\
Semimajor axis in stellar radii  (a/R$_*$)   &  $\mathcal{U}$(10,200)  & 25.21$^{+3.51}_{-2.24}$  &  24.40$^{+2.21}_{-2.09}$ \\
Inclination (i) [deg]   &  $\mathcal{U}$(0,90) & 88.62$^{+0.57}_{-0.35}$ &  88.95$^{+0.64}_{-0.57}$ \\
Eccentricity (e) &  $\mathcal{U}$(0,1)  &  - &  0.18$^{+0.07}_{-0.06}$ \\
Argument of periastron ($\omega$) [rad] &  $\mathcal{U}$(0,2$\Pi$) & - &  4.25$^{+0.26}_{-0.31}$ \\
Semi-amplitude (K) [m.s$^{-1}$] & $\mathcal{U}$(0,10)  & 2.15$^{+0.19}_{-0.19}$  & 2.24$^{+0.18}_{-0.18}$ \\
\hline
\multicolumn{4}{c}{planet c}\\
Inferior conjunction (T$_0$) [BJD - 2457000]   &  $\mathcal{U}$(1421,1426)  & 1425.078344$^{+0.000574}_{-0.000544}$ &  1425.078182$^{+0.000472}_{-0.000450}$\\
Orbital period (P) [days]   &  $\mathcal{U}$(2.5,3.5) & 3.123898$^{+0.000003}_{-0.000003}$  &  3.123898$^{+0.000003}_{-0.000003}$\\
Radius in stellar radii (R$_p$/R$_*$)   &  $\mathcal{U}$(0.02,0.1) &  0.05313$^{+0.02038}_{-0.00924}$  &  0.05792$^{+0.01974}_{-0.01188}$\\
Inclination (i) [deg]   &  $\mathcal{U}$(0,90) & 86.79$^{+0.46}_{-0.39}$ &  86.63$^{+0.39}_{-0.47}$ \\
Eccentricity (e) &  $\mathcal{U}$(0,1)  &  - &  0.05$^{+0.05}_{-0.03}$ \\
Argument of periastron ($\omega$) [rad] &  $\mathcal{U}$(0,2$\Pi$) & - &  2.84$^{+2.38}_{-1.89}$ \\
Semi-amplitude (K) [m.s$^{-1}$] & $\mathcal{U}$(0,10)  & 1.11$^{+0.20}_{-0.20}$  & 1.13$^{+0.18}_{-0.18}$ \\
\hline
\multicolumn{4}{c}{planet d}\\
Inferior conjunction (T$_0$) [BJD - 2457000]   &  $\mathcal{U}$(1421,1450)  & 1441.07$^{+1.80}_{-1.21}$ &  1440.09$^{+1.53}_{-1.39}$\\
Orbital period (P) [days]   &  $\mathcal{U}$(23,25) & 24.30$^{+0.03}_{-0.08}$  &  24.29$^{+0.03}_{-0.05}$\\
Eccentricity (e) &  $\mathcal{U}$(0,1)  &  - &  0.26$^{+0.14}_{-0.13}$ \\
Argument of periastron ($\omega$) [rad] &  $\mathcal{U}$(0,2$\Pi$) & - &  0.88$^{+0.87}_{-0.59}$ \\
Semi-amplitude (K) [m.s$^{-1}$] & $\mathcal{U}$(0,10)  & 1.53$^{+0.40}_{-0.41}$  & 1.60$^{+0.36}_{-0.37}$ \\
\hline
\hline
\multicolumn{4}{c}{Light curve model}\\
Linear limb-darkening coefficient (u$_1$) & $\mathcal{U}$(0,1) & 0.15$^{+0.15}_{-0.11}$ & 0.15$^{+0.13}_{-0.10}$ \\
Quadratic limb-darkening coefficient (u$_2$) & $\mathcal{U}$(0,1) & 0.25$^{+0.24}_{-0.17}$ & 0.29$^{+0.23}_{-0.19}$ \\
\multicolumn{4}{c}{RV model}\\
Systemic RV ESPRESSO ($\gamma_{ESP}$) [m.s$^{-1}$] & $\mathcal{U}$(-50000,50000) & -5404.26$^{+1.79}_{-1.76}$ & -5404.17$^{+1.59}_{-1.56}$ \\
Systemic RV HARPS ($\gamma_{HARPS}$) [m.s$^{-1}$] & $\mathcal{U}$(-50000,50000) & -5442.55$^{+1.74}_{-1.79}$ & -5442.58$^{+1.58}_{-1.61}$ \\
RV slope ($\dot{\gamma}$) [m.s$^{-1}$.day$^{-1}$]  & $\mathcal{U}$(-10,10)  & -0.012149$^{+0.000677}_{-0.000686}$ & -0.012186$^{+0.000590}_{-0.000618}$ \\
RV jitter ESPRESSO ($\sigma_{ESP}$) [m.s$^{-1}$] & $\mathcal{U}$(0,1.5) & 0.93$^{+0.17}_{-0.15}$ & 0.88$^{+0.15}_{-0.13}$ \\
RV jitter HARPS ($\sigma_{HARPS}$) [m.s$^{-1}$]  & $\mathcal{U}$(0.5,3) & 1.26$^{+0.80}_{-0.52}$ & 1.13$^{+0.72}_{-0.45}$ \\
\hline
\multicolumn{4}{c}{GP Hyperparameters }\\
Amplitude RV GP [m.s$^{-1}$] & $\mathcal{U}$(0,20) & 10.43$^{+3.82}_{-2.93}$ & 9.06$^{+3.05}_{-2.32}$ \\

Period RV GP$^{1}$ (P$_{rot}$) [days] & $\mathcal{U}$(0,200) & \multicolumn{2}{c}{72} \\

Length RV GP$^{1}$ (L$_{act}$) [days] & $\mathcal{U}$(0,200) & \multicolumn{2}{c}{72} \\

Amplitude LC GP$^{2}$ log[] & $\mathcal{U}$(-20,20) & \multicolumn{2}{c}{-11.10$^{+0.24}_{-0.19}$}  \\
Period LC GP$^{2}$ (P) [days] & $\mathcal{U}$(1,2) & \multicolumn{2}{c}{1.406$^{+0.014}_{-0.014}$}  \\
Length LC GP$^{2}$ (L) log[days] & $\mathcal{U}$(0,10) & \multicolumn{2}{c}{1.75$^{+0.24}_{-0.20}$}  \\
\hline
\hline
\multicolumn{4}{l}{$^1$Parameter fixed}\\
\multicolumn{4}{l}{$^2$Parameter retrieved with the binned light curves}


\end{tabular}
\end{center}
\end{table*}

\clearpage

\begin{acknowledgements}
We are grateful for a constructive review from Sarah Blunt, which allowed us to deepen our analysis.The authors acknowledge the ESPRESSO project team for its effort and dedication in building the ESPRESSO instrument. This work has been carried out in the frame of the National Centre for Competence in Research `PlanetS' supported by the Swiss National Science Foundation (SNSF). B.L.,A.D. acknowledge the financial support of the SNSF. A.D. acknowledges support from the European Research Council (ERC) under the European Union's Horizon 2020 research and innovation programme (project {\sc Four Aces}, grant agreement No. 724427). This work has made use of data from the European Space Agency (ESA) mission {\it Gaia} (\url{https://www.cosmos.esa.int/gaia}), processed by the {\it Gaia} Data Processing and Analysis Consortium (DPAC, \url{https://www.cosmos.esa.int/web/gaia/dpac/consortium}). Funding for the DPAC has been provided by national institutions, in particular the institutions participating in the {\it Gaia} Multilateral Agreement. ASM, JIGH, CAP and RR acknowledge financial support from the Spanish Ministry of Science and Innovation (MICINN) project PID2020-117493GB-I00. ASM, JIGH and RR also acknowledge financial support from the Government of the Canary Islands project ProID2020010129.V.A. acknowledges the support from FCT through Investigador FCT contract nr.  IF/00650/2015/CP1273/CT0001. JIGH also acknowledges financial support from the Spanish MICINN under 2013 Ram\'on y Cajal program RYC-2013-14875. NJN acknowledges support form the following projects: UIDB/04434/2020 \& UIDP/04434/2020, CERN/FIS-PAR/0037/2019, PTDC/FIS-OUT/29048/2017, COMPETE2020: POCI-01-0145-FEDER-028987 \& FCT: PTDC/FIS-AST/28987/2017. A.S.M. acknowledges financial support from the Spanish Ministry of Science and Innovation (MICINN) under the 2018 Juan de la Cierva Programme IJC2018-035229-I. IAC: This work is partly financed by the Spanish Ministry of Economics and Competitiveness through grants PGC2018-098153-B-C31.
\end{acknowledgements}

\bibliographystyle{aa} 
\bibliography{biblio_Oct2020} 

\begin{appendix}
\section{Appendix}
\begin{figure*}
\centering
\includegraphics[width= \columnwidth]{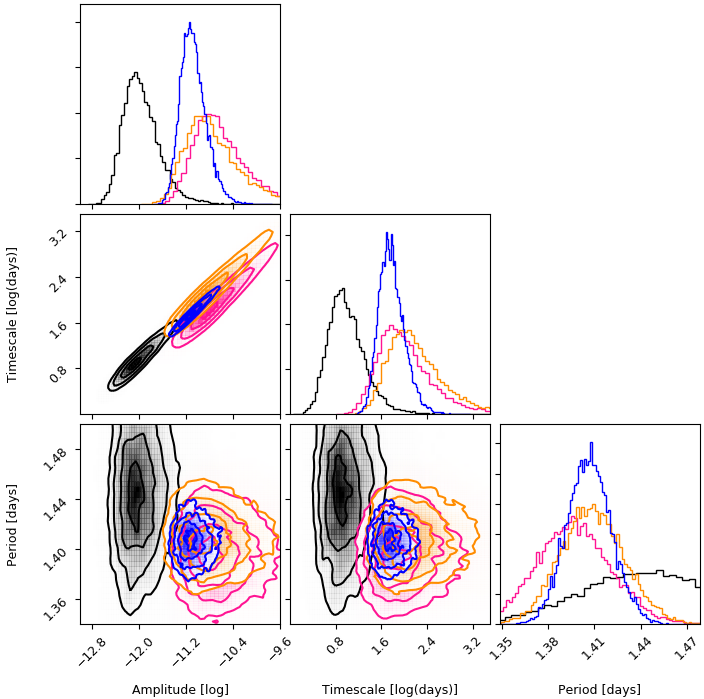}
\caption{\label{post_hyperparam_1pt4}Posterior distribution of the GP hyperparameters for the 1.4-day signal in the TESS light curves from sector 4 (black), sector 31 with 2-minute cadence (orange), sector 31 with 20-second cadence (pink), and the combined light-curves dataset (blue). Priors are uniform between [-20,20] m.s$^{-1}$ for the logarithm of the amplitude, [0,10] days for the logarithm of the length, and [1,2] days for the period.}
\end{figure*}
\begin{figure*}
\centering
\includegraphics[width= 2\columnwidth]{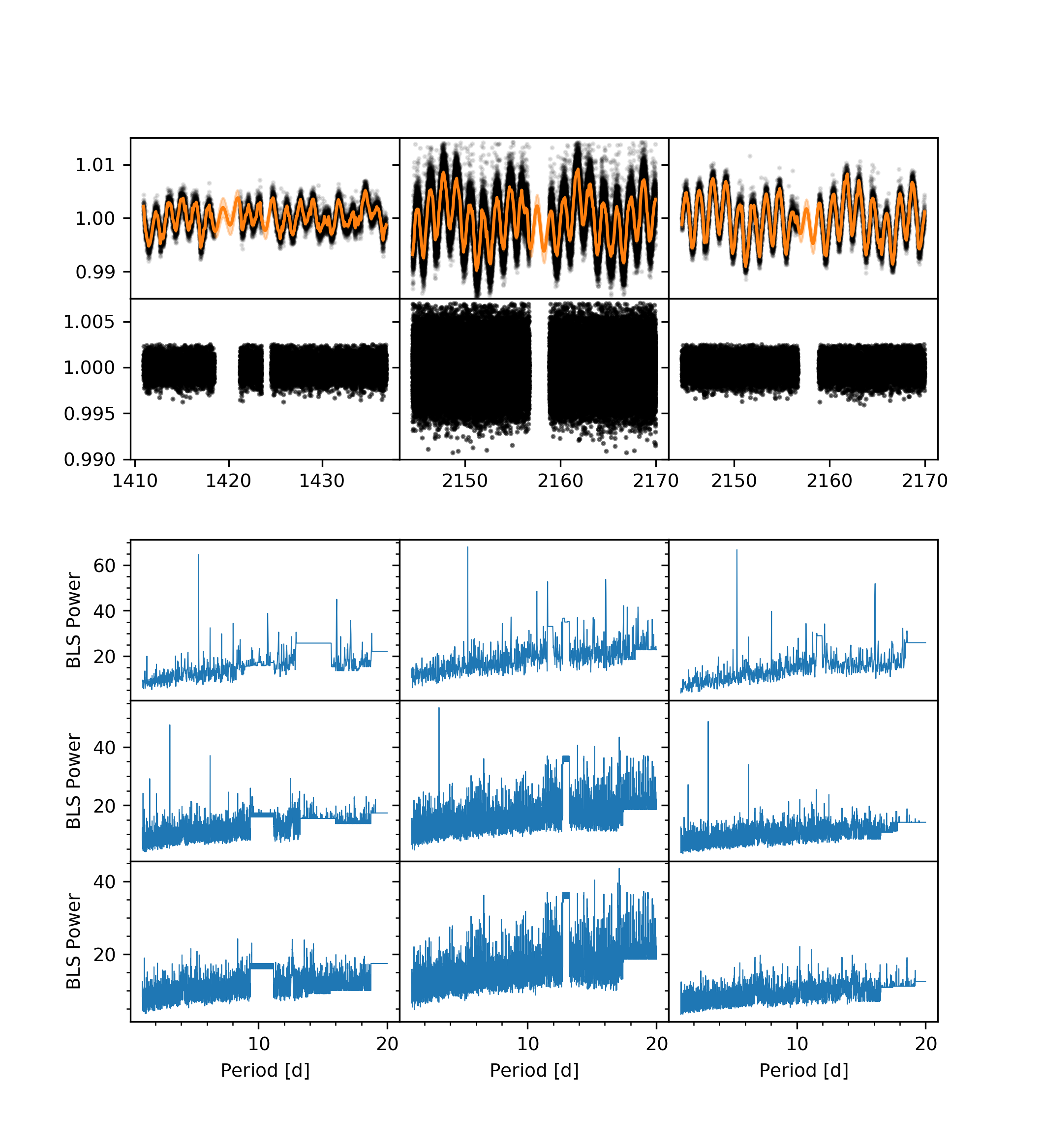}
\caption{\label{iterative_period_lc}Identification of planet candidates for TESS light curves in sector 4 (first column), sector 31 with 20-second cadence (second column), and sector 31 with 2-minute cadence (third column).  The first row shows the PDCSAP light curves with the GP fit (see section \ref{identification_transit}). The second row shows the detrended light curves. The last panels show the iterative periodogram analysis. The detected signals are at 5.35 days (third row) and 3.12 days (fourth row). The remaining time series has no signal (last row).  }
\end{figure*}

\clearpage
\begin{figure*}
\centering
\includegraphics[width= 2\columnwidth]{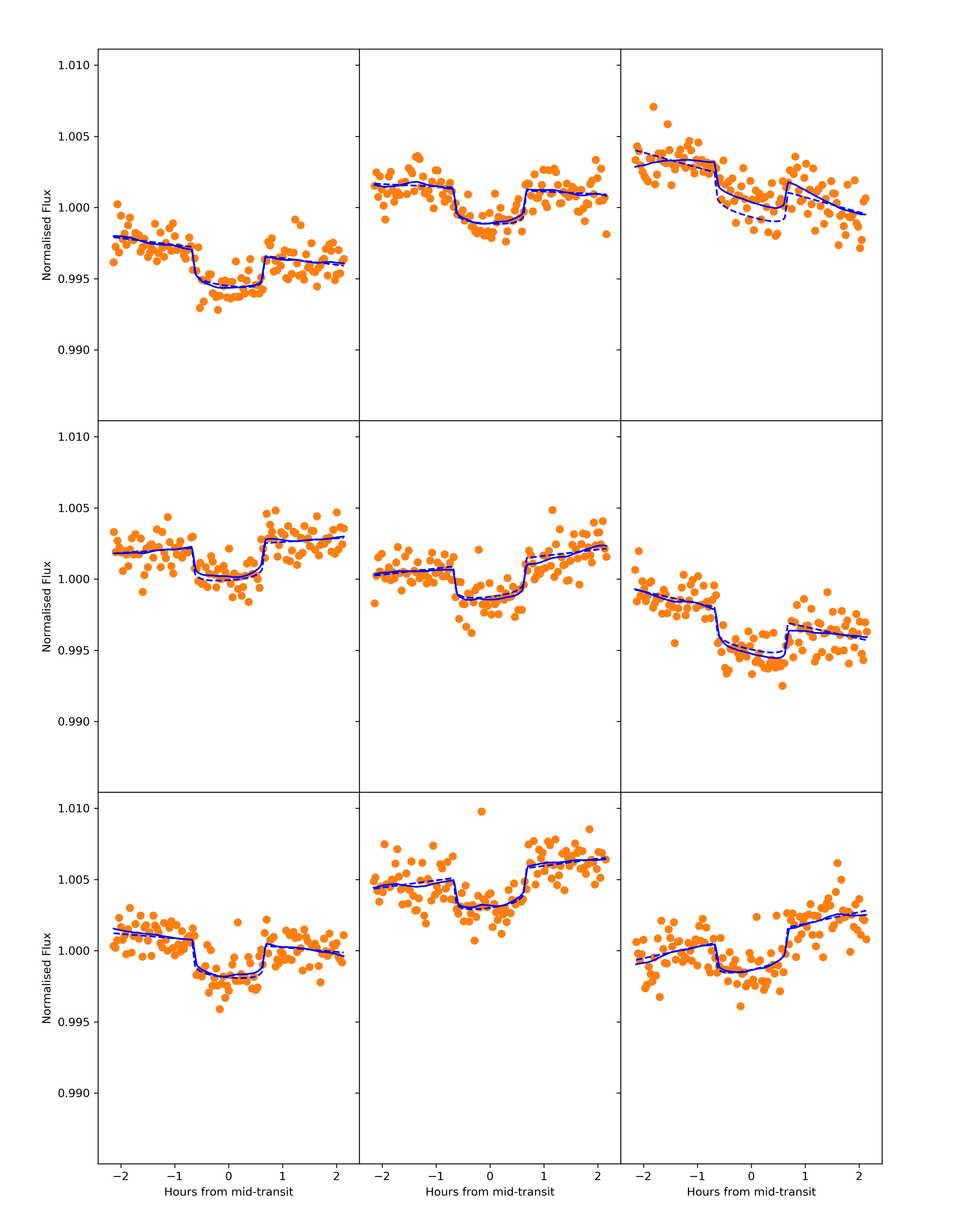}
\caption{\label{gp_vs_poly_fit_b} Light curves centred around each transit of planet b. The plain blue line indicates the best fit from the Bayesian inference done on the combined light curves of both sectors with the joint GP-transit model. The dashed blue line shows the best fit from the Bayesian inference done with a circular transit model on the \textit{\textup{reduced dataset}} (two hours before and after mid-transits) detrended with a linear model. }
\end{figure*}
\begin{figure*}
\centering
\includegraphics[width= 1.3\columnwidth]{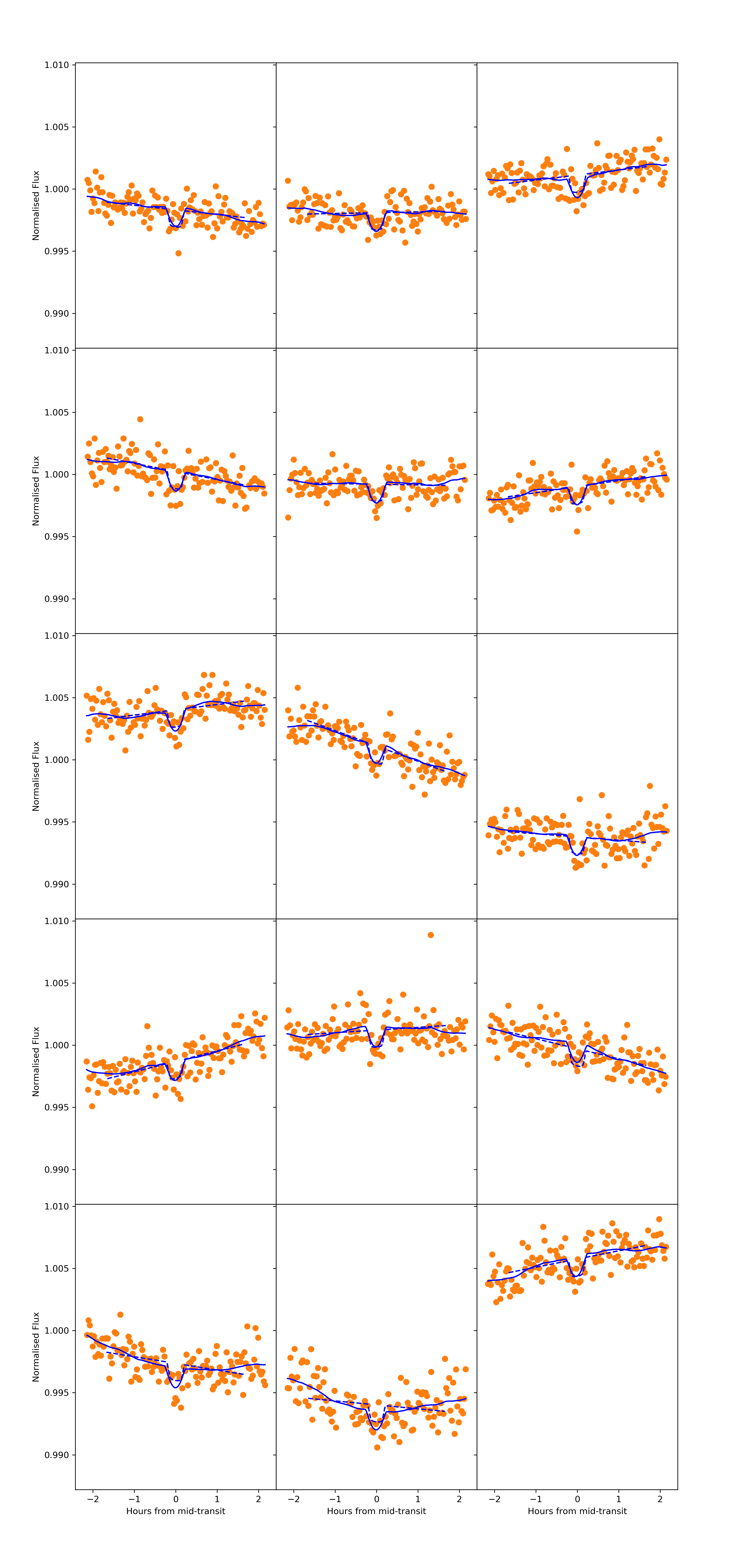}
\caption{\label{gp_vs_poly_fit_c}Same as Figure \ref{gp_vs_poly_fit_b}, but for planet c}
\end{figure*}


\begin{figure*}
\centering
\includegraphics[width=2\columnwidth]{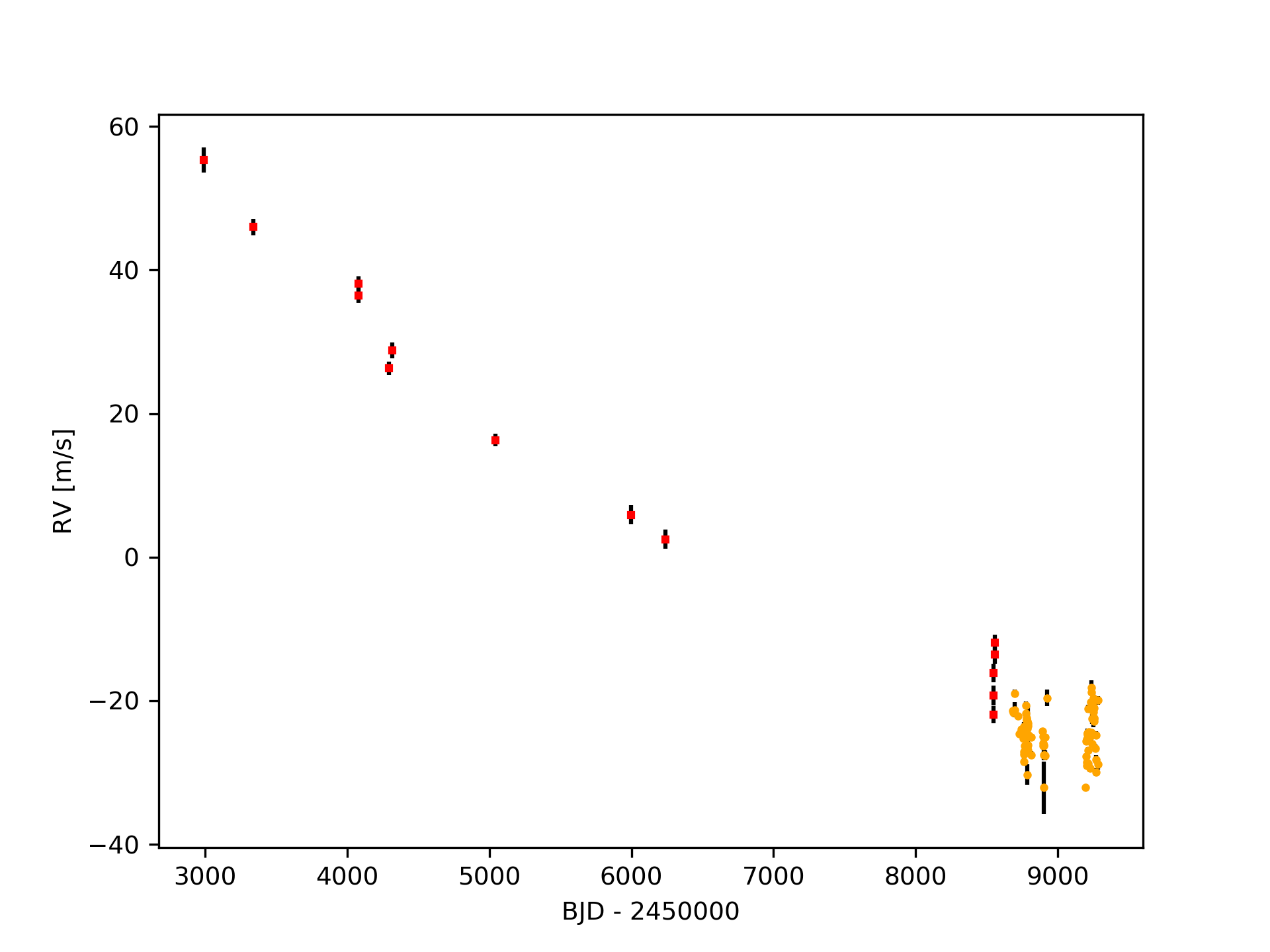}
\caption{\label{rv_all} Radial velocity measurements of LTT 1445A with HARPS (red) and ESPRESSO (orange). There is a clear long-term drift due to the interaction of the star with the BC binary of the triple system.
}
\end{figure*}

\begin{figure*}
\centering
\includegraphics[width=2\columnwidth]{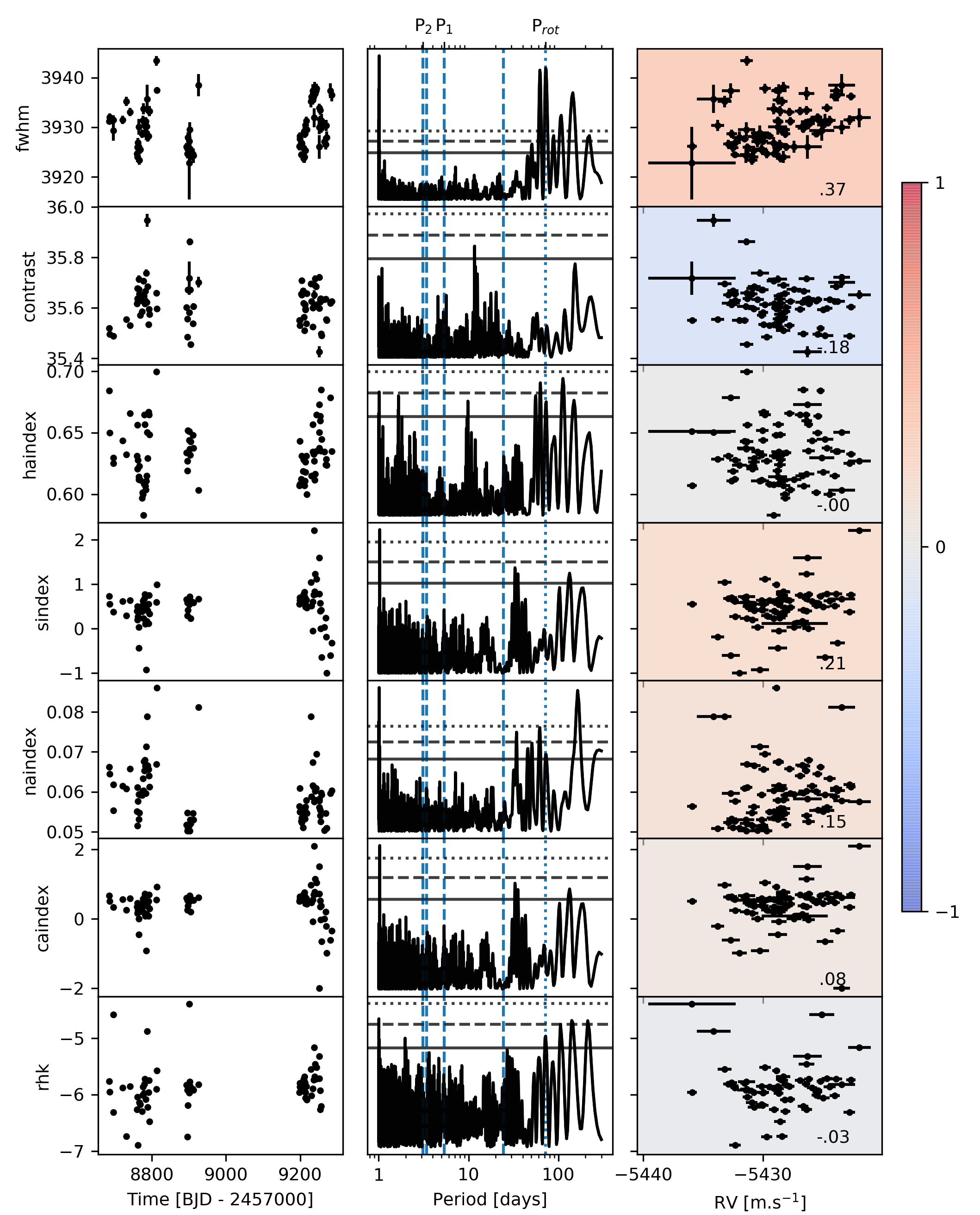}
\caption{\label{periodograms_all} Time series, Lomb-Scargle periodograms, and correlation with radial velocities for each activity index of the ESPRESSO dataset for LTT 1445A.
For the middle column, the dotted, dashed, and plain horizontal lines indicate the levels for a false-alarm probability of 0.1\%, 1\%, and 10\%,   respectively.  
For the last column, the Pearson correlation coefficient is computed for each pair (activity index and RV). It is indicated as a number and with a face colour in each subplot. A coefficient of 1 (-1) indicates that the activity indey is correlated (anti-correlated) with the radial velocities. A coefficient of 0 indicates no correlation. 
}
\end{figure*}

\begin{figure*}
\centering
\includegraphics[width=1.75\columnwidth]{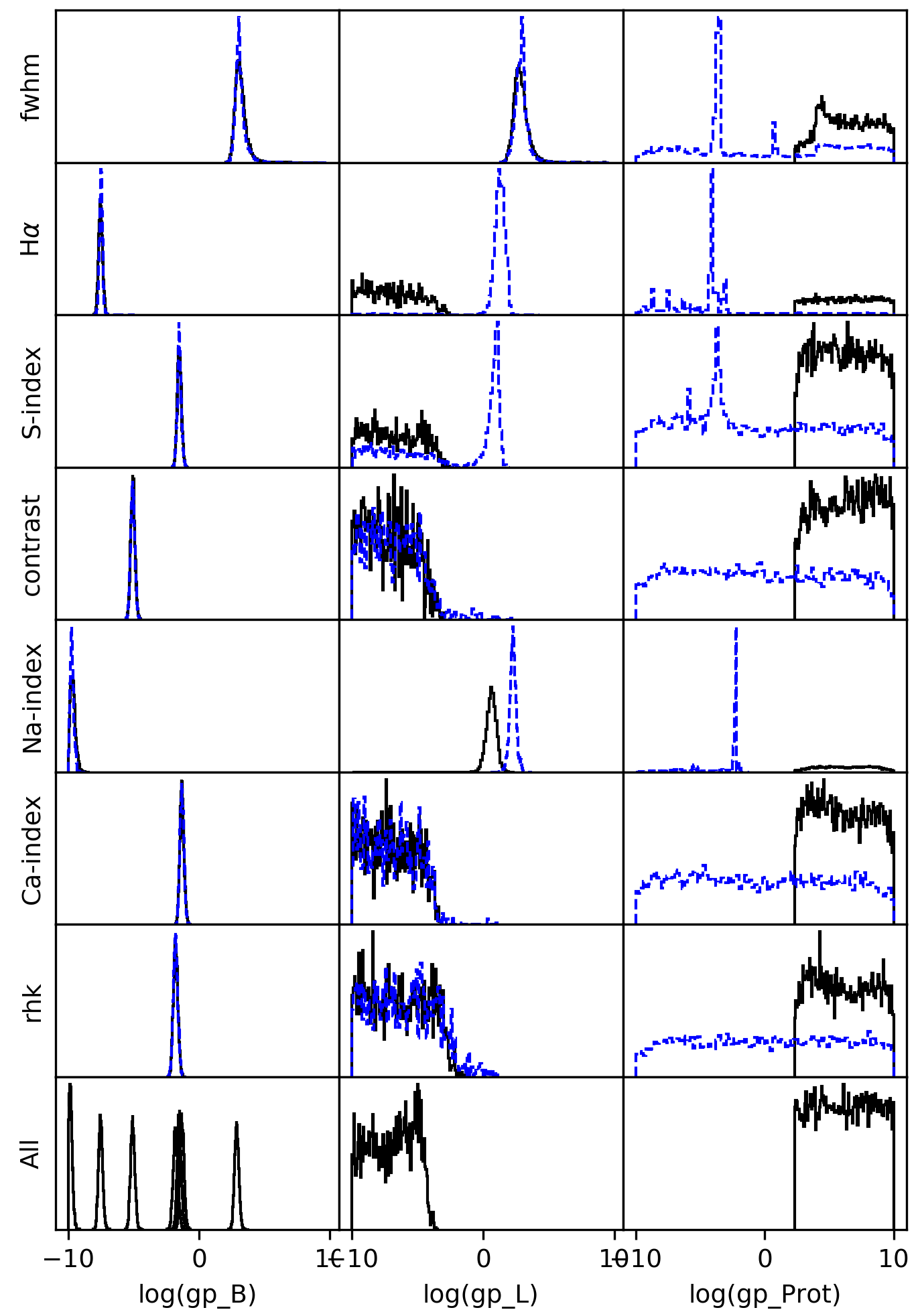}
\caption{\label{Ap_TSPcorr} Posterior distributions for the hyperparameters of the GP for the inference on the seven activity indicators of ESPRESSO (first seven rows) independently and for the combined inference with the seven indices (last row). The dotted blue histograms are from the broad uniform priors, and the black histograms are from the inference with a priorlonger than 35 days for the P$_{rot}$ parameter.}
\end{figure*}

\begin{figure*}
\centering
\includegraphics[width=2\columnwidth]{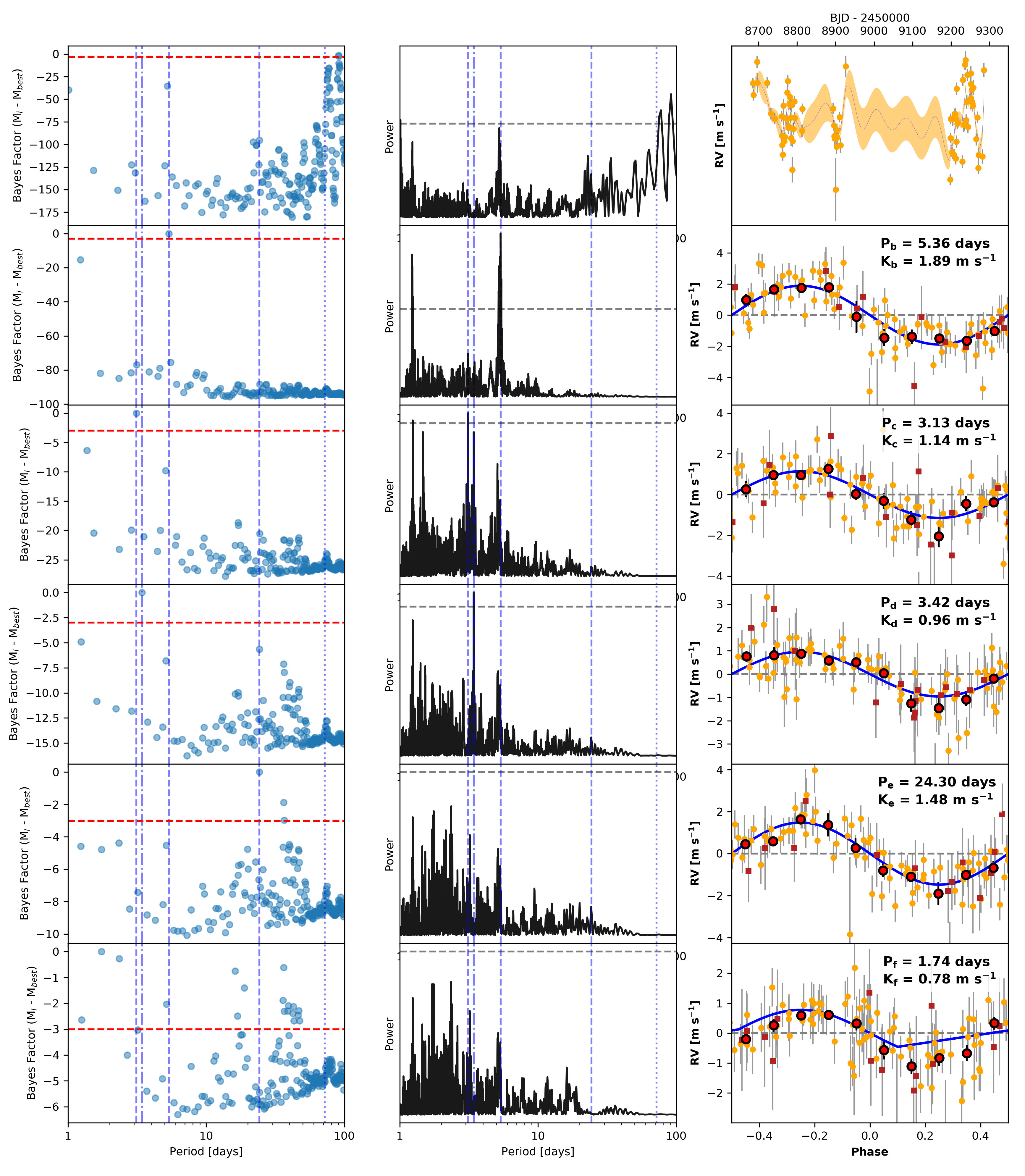}
\caption{\label{blindRV_App} Blind search for Keplerian signals in the radial velocity dataset. The period parameter space is subdivided into bins of 0.5 days. An iteration corresponds to a Bayesian inference for each bin with a Keplerian model (first row) or a joint GP-Keplerian model with one to five planets (rows 2 to 6). The first column shows the logarithm of the Bayes factor of each model/period bin with the best model/period bin of the iteration. The second column shows the Lomb-Scargle periodogram computed with the residue of the best model from the previous iteration (except for the first row, which is the periodogram from the data).  The last column shows the best fit to the data from the best model/period of the iteration, except for the first row, which shows the GP fit to the ESPRESSO dataset. The dashed vertical lines for columns 1 and 2 indicate the three Keplerian signals at 5.4, 3.1, and 24.3 days, the dotted vertical line indicates the 72-day period associated with the stellar rotation, and the dotted-dashed vertical line indicates the polluting signal at 3.4 days. The horizontal dashed line in column 1 indicates the moderate threshold for the model selection. The horizontal dashed line in column 2 indicates the level for a false-alarm probability of  1\%. 
}
\end{figure*}

\begin{figure}
\centering
\includegraphics[width=\columnwidth]{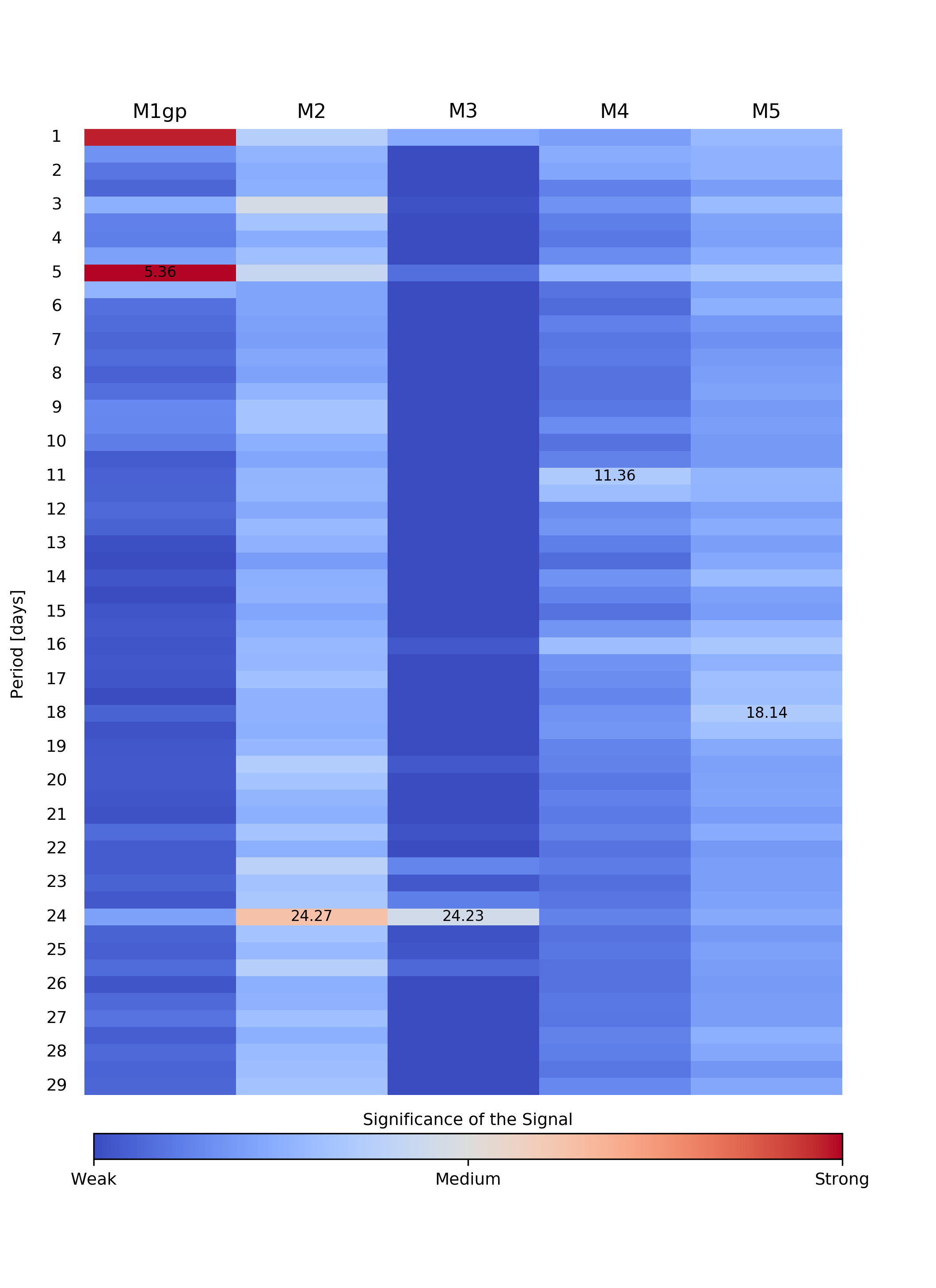}
\caption{\label{hotmap_blindRV_1pt4} Same as Figure \ref{hotmap_blindRV}, but with the 1.4 days GP.
}
\end{figure}

\clearpage

\begin{table*}
\tiny
\begin{center}
\caption{\label{tadder} Model ladder }
\begin{tabular}{cccccccccccc}
\hline
\hline
model name &  number of   & number of  & Jitter & Slope & Curvature & GP & LogZ & Bayes & semi-amplitude & semi-amplitude & semi-amplitude  \\
 &  parameter  & planets &  &  &  &  &  &  & planet b & planet c & planet d   \\
\hline
M4ccecJLgp1 &  20  & 4 & yes & yes & no & 72 & -246.4 & 0.9 & 2.2$^{+0.2}_{-0.2}$ & 1.1$^{+0.2}_{-0.2}$ & 1.5$^{+0.4}_{-0.4}$ \\ 
M4eeeeJLgp1 &  26  & 4 & yes & yes & no & 72 & -247.3 & 0.2 & 2.2$^{+0.2}_{-0.2}$ & 1.0$^{+0.2}_{-0.2}$ & 1.6$^{+0.4}_{-0.4}$ \\ 
M5cceccJLgp1 &  23  & 5 & yes & yes & no & 72 & -247.4 & 0.2 & 2.2$^{+0.2}_{-0.2}$ & 1.1$^{+0.2}_{-0.2}$ & 1.6$^{+0.4}_{-0.4}$ \\ 
M4ccccJLgp1 &  18  & 4 & yes & yes & no & 72 & -247.7 & 0.7 & 2.2$^{+0.2}_{-0.2}$ & 1.1$^{+0.2}_{-0.2}$ & 1.6$^{+0.4}_{-0.4}$ \\ 
M5cccccJLgp1 &  21  & 5 & yes & yes & no & 72 & -248.4 & 0.0 & 2.2$^{+0.2}_{-0.2}$ & 1.1$^{+0.2}_{-0.2}$ & 1.6$^{+0.4}_{-0.4}$ \\ 
M4eeeeGLgp1 &  24  & 4 & no & yes & no & 72 & -248.4 & 0.1 & 2.2$^{+0.2}_{-0.2}$ & 1.0$^{+0.2}_{-0.2}$ & 1.5$^{+0.5}_{-0.4}$ \\ 
M3cceJLgp1 &  17  & 3 & yes & yes & no & 72 & -248.5 & 0.1 & 2.2$^{+0.2}_{-0.2}$ & 1.0$^{+0.2}_{-0.2}$ & - \\ 
M4ccecGLgp1 &  18  & 4 & no & yes & no & 72 & -248.5 & 0.9 & 2.2$^{+0.2}_{-0.2}$ & 1.1$^{+0.2}_{-0.2}$ & 1.3$^{+0.5}_{-0.4}$ \\ 
M4ccecJLgp1$^{1.4}$ &  20  & 4 & yes & yes & no & 72 & -249.4 & 0.0 & 2.1$^{+0.2}_{-0.2}$ & 1.2$^{+0.2}_{-0.2}$ & 1.7$^{+0.4}_{-0.4}$ \\ 
M3eeeJLgp1 &  21  & 3 & yes & yes & no & 72 & -249.4 & 0.4 & 2.2$^{+0.2}_{-0.2}$ & 1.0$^{+0.2}_{-0.2}$ & - \\ 
M3cceGLgp1 &  15  & 3 & no & yes & no & 72 & -249.9 & 0.1 & 2.2$^{+0.2}_{-0.2}$ & 1.0$^{+0.2}_{-0.2}$ & - \\ 
M3cccJLgp1 &  15  & 3 & yes & yes & no & 72 & -250.0 & 0.4 & 2.2$^{+0.2}_{-0.2}$ & 1.1$^{+0.2}_{-0.2}$ & - \\ 
M4eeeeJLgp1$^{1.4}$ &  26  & 4 & yes & yes & no & 72 & -250.3 & 0.1 & 2.2$^{+0.2}_{-0.2}$ & 1.2$^{+0.2}_{-0.2}$ & 1.8$^{+0.5}_{-0.4}$ \\ 
M4ccccJLgp1$^{1.4}$ &  18  & 4 & yes & yes & no & 72 & -250.5 & 0.1 & 2.1$^{+0.2}_{-0.2}$ & 1.2$^{+0.2}_{-0.2}$ & 1.7$^{+0.5}_{-0.4}$ \\ 
M3eeeGLgp1 &  19  & 3 & no & yes & no & 72 & -250.6 & 0.5 & 2.2$^{+0.2}_{-0.2}$ & 1.0$^{+0.2}_{-0.2}$ & - \\ 
M3cceJLgp1$^{1.4}$ &  17  & 3 & yes & yes & no & 72 & -251.1 & 0.2 & 2.1$^{+0.2}_{-0.2}$ & 1.1$^{+0.2}_{-0.2}$ & - \\ 
M2ccJLgp1 &  12  & 2 & yes & yes & no & 72 & -251.3 & 0.7 & 2.1$^{+0.2}_{-0.2}$ & 1.1$^{+0.2}_{-0.2}$ & - \\ 
M3eeeJLgp1$^{1.4}$ &  21  & 3 & yes & yes & no & 72 & -251.9 & 0.0 & 2.1$^{+0.2}_{-0.2}$ & 1.1$^{+0.2}_{-0.2}$ & - \\ 
M4ccccGLgp2 &  16  & 4 & no & yes & no & 1.4 & -252.0 & 0.1 & 2.2$^{+0.2}_{-0.2}$ & 1.3$^{+0.3}_{-0.3}$ & 1.9$^{+0.3}_{-0.3}$ \\ 
M2eeJLgp1 &  16  & 2 & yes & yes & no & 72 & -252.1 & 0.0 & 2.1$^{+0.2}_{-0.2}$ & 1.1$^{+0.3}_{-0.3}$ & - \\ 
M3eeeGLgp1$^{1.4}$ &  19  & 3 & no & yes & no & 72 & -252.1 & 0.1 & 2.2$^{+0.1}_{-0.2}$ & 1.1$^{+0.1}_{-0.2}$ & - \\ 
M3cccGLgp2 &  13  & 3 & no & yes & no & 1.4 & -252.2 & 0.2 & 2.2$^{+0.2}_{-0.2}$ & 1.2$^{+0.3}_{-0.3}$ & 1.8$^{+0.4}_{-0.4}$ \\ 
M4ccccGLgp1 &  16  & 4 & no & yes & no & 72 & -252.4 & 0.1 & 2.2$^{+0.2}_{-0.2}$ & 1.1$^{+0.2}_{-0.2}$ & 1.3$^{+0.5}_{-0.4}$ \\ 
M3cccJLgp1$^{1.4}$ &  15  & 3 & yes & yes & no & 72 & -252.4 & 0.0 & 2.1$^{+0.2}_{-0.2}$ & 1.1$^{+0.2}_{-0.2}$ & - \\ 
M3cccJLgp2 &  15  & 3 & yes & yes & no & 1.4 & -252.4 & 0.1 & 2.2$^{+0.2}_{-0.2}$ & 1.2$^{+0.3}_{-0.3}$ & 1.8$^{+0.4}_{-0.4}$ \\ 
M4ccccJLgp2 &  18  & 4 & yes & yes & no & 1.4 & -252.5 & 0.4 & 2.2$^{+0.2}_{-0.2}$ & 1.3$^{+0.3}_{-0.3}$ & 1.8$^{+0.3}_{-0.3}$ \\ 
M4ccecGLgp1$^{1.4}$ &  18  & 4 & no & yes & no & 72 & -252.9 & 0.6 & 2.1$^{+0.1}_{-0.1}$ & 1.2$^{+0.2}_{-0.2}$ & 1.2$^{+0.4}_{-0.4}$ \\ 
M4eeeeGLgp2 &  24  & 4 & no & yes & no & 1.4 & -253.4 & 0.2 & 2.3$^{+0.2}_{-0.2}$ & 1.2$^{+0.3}_{-0.3}$ & 1.9$^{+0.3}_{-0.3}$ \\ 
M3eeeGLgp2 &  19  & 3 & no & yes & no & 1.4 & -253.6 & 0.2 & 2.3$^{+0.2}_{-0.2}$ & 1.2$^{+0.3}_{-0.3}$ & 1.8$^{+0.4}_{-0.4}$ \\

\hline
\end{tabular}
\end{center}
\end{table*}

\begin{table*}
\tiny
\begin{center}
\caption{\label{ladder_2} Model ladder }
\begin{tabular}{cccccccccccc}
\hline
\hline
model name &  number of   & number of  & Jitter & Slope & Curvature & GP & LogZ & Bayes & semi-amplitude & semi-amplitude & semi-amplitude  \\
 &  parameter  & planets &  &  &  &  &  &  & planet b & planet c & planet d   \\
\hline
M3cccGLgp1 &  13  & 3 & no & yes & no & 72 & -253.8 & 0.0 & 2.2$^{+0.2}_{-0.2}$ & 1.1$^{+0.2}_{-0.2}$ & - \\ 
M3eeeJLgp2 &  21  & 3 & yes & yes & no & 1.4 & -253.8 & 0.1 & 2.3$^{+0.2}_{-0.2}$ & 1.2$^{+0.3}_{-0.3}$ & 1.8$^{+0.4}_{-0.4}$ \\ 
M4eeeeJLgp2 &  26  & 4 & yes & yes & no & 1.4 & -254.0 & 0.2 & 2.3$^{+0.2}_{-0.2}$ & 1.2$^{+0.3}_{-0.3}$ & 1.9$^{+0.3}_{-0.3}$ \\ 
M3cceGLgp1$^{1.4}$ &  15  & 3 & no & yes & no & 72 & -254.2 & 2.4 & 2.0$^{+0.2}_{-0.2}$ & 1.2$^{+0.2}_{-0.2}$ & - \\ 
M4ccecJCgp1 &  21  & 4 & yes & yes & yes & 72 & -256.6 & 0.4 & 2.1$^{+0.2}_{-0.2}$ & 1.1$^{+0.2}_{-0.2}$ & 1.5$^{+0.4}_{-0.4}$ \\ 
M4ccccGLgp1$^{1.4}$ &  16  & 4 & no & yes & no & 72 & -257.0 & 0.0 & 2.2$^{+0.1}_{-0.1}$ & 1.1$^{+0.2}_{-0.2}$ & 1.4$^{+0.4}_{-0.4}$ \\ 
M4eeeeJCgp1 &  27  & 4 & yes & yes & yes & 72 & -257.1 & 0.6 & 2.1$^{+0.2}_{-0.2}$ & 1.0$^{+0.2}_{-0.2}$ & 1.5$^{+0.4}_{-0.4}$ \\ 
M5cceccJCgp1 &  24  & 5 & yes & yes & yes & 72 & -257.6 & 0.0 & 2.1$^{+0.2}_{-0.2}$ & 1.1$^{+0.2}_{-0.2}$ & 1.5$^{+0.3}_{-0.3}$ \\ 
M4ccccJCgp1 &  19  & 4 & yes & yes & yes & 72 & -257.7 & 0.5 & 2.1$^{+0.2}_{-0.2}$ & 1.1$^{+0.2}_{-0.2}$ & 1.5$^{+0.4}_{-0.4}$ \\ 
M1cJLgp1 &  9  & 1 & yes & yes & no & 72 & -258.2 & 0.2 & 2.0$^{+0.2}_{-0.3}$ & - & - \\ 
M1eJLgp1 &  11  & 1 & yes & yes & no & 72 & -258.3 & 0.0 & 2.0$^{+0.3}_{-0.3}$ & - & - \\ 
M5cccccJCgp1 &  22  & 5 & yes & yes & yes & 72 & -258.3 & 0.0 & 2.1$^{+0.2}_{-0.2}$ & 1.1$^{+0.2}_{-0.2}$ & 1.6$^{+0.3}_{-0.3}$ \\ 
M3cccGLgp1$^{1.4}$ &  13  & 3 & no & yes & no & 72 & -258.3 & 0.9 & 2.2$^{+0.1}_{-0.1}$ & 1.0$^{+0.2}_{-0.2}$ & - \\ 
M4ccecJCgp1$^{1.4}$ &  21  & 4 & yes & yes & yes & 72 & -259.2 & 0.1 & 2.0$^{+0.2}_{-0.2}$ & 1.1$^{+0.2}_{-0.2}$ & 1.6$^{+0.4}_{-0.3}$ \\ 
M4eeeeJCgp1$^{1.4}$ &  27  & 4 & yes & yes & yes & 72 & -259.4 & 0.3 & 2.1$^{+0.2}_{-0.2}$ & 1.1$^{+0.2}_{-0.2}$ & 1.8$^{+0.4}_{-0.4}$ \\ 
M4eeeeGCgp1 &  25  & 4 & no & yes & yes & 72 & -259.6 & 0.0 & 2.2$^{+0.2}_{-0.2}$ & 1.0$^{+0.2}_{-0.2}$ & 1.4$^{+0.4}_{-0.4}$ \\ 
M3cceJCgp1 &  18  & 3 & yes & yes & yes & 72 & -259.6 & 0.4 & 2.1$^{+0.2}_{-0.2}$ & 1.0$^{+0.2}_{-0.2}$ & - \\ 
M2ccGLgp2 &  10  & 2 & no & yes & no & 1.4 & -260.1 & 0.2 & 2.1$^{+0.2}_{-0.2}$ & 1.2$^{+0.3}_{-0.3}$ & - \\ 
M2ccJLgp2 &  12  & 2 & yes & yes & no & 1.4 & -260.2 & 0.0 & 2.1$^{+0.2}_{-0.3}$ & 1.2$^{+0.4}_{-0.4}$ & - \\ 
M3eeeJCgp1 &  22  & 3 & yes & yes & yes & 72 & -260.3 & 0.2 & 2.2$^{+0.2}_{-0.2}$ & 1.0$^{+0.2}_{-0.2}$ & - \\ 
M4ccccJCgp1$^{1.4}$ &  19  & 4 & yes & yes & yes & 72 & -260.4 & 0.1 & 2.0$^{+0.2}_{-0.2}$ & 1.1$^{+0.2}_{-0.2}$ & 1.6$^{+0.4}_{-0.3}$ \\ 
M4ccecGCgp1 &  19  & 4 & no & yes & yes & 72 & -260.5 & 0.4 & 2.2$^{+0.2}_{-0.2}$ & 1.1$^{+0.2}_{-0.2}$ & 1.3$^{+0.4}_{-0.4}$ \\ 
M4ccccGCgp2 &  17  & 4 & no & yes & yes & 1.4 & -260.9 & 0.1 & 2.2$^{+0.2}_{-0.2}$ & 1.2$^{+0.3}_{-0.3}$ & 1.7$^{+0.3}_{-0.3}$ \\ 
M2eeGLgp2 &  14  & 2 & no & yes & no & 1.4 & -261.0 & 0.1 & 2.2$^{+0.3}_{-0.3}$ & 1.2$^{+0.4}_{-0.5}$ & - \\ 
M3cccJCgp1 &  16  & 3 & yes & yes & yes & 72 & -261.1 & 0.2 & 2.1$^{+0.2}_{-0.2}$ & 1.0$^{+0.2}_{-0.2}$ & - \\ 
M2eeJLgp2 &  16  & 2 & yes & yes & no & 1.4 & -261.3 & 0.3 & 2.2$^{+0.3}_{-0.3}$ & 1.2$^{+0.4}_{-0.5}$ & - \\ 
M4ccccJCgp2 &  19  & 4 & yes & yes & yes & 1.4 & -261.6 & 0.0 & 2.1$^{+0.2}_{-0.2}$ & 1.2$^{+0.3}_{-0.3}$ & 1.7$^{+0.3}_{-0.3}$ \\ 
M3cccJCgp2 &  16  & 3 & yes & yes & yes & 1.4 & -261.6 & 0.1 & 2.1$^{+0.2}_{-0.2}$ & 1.1$^{+0.3}_{-0.3}$ & 1.6$^{+0.3}_{-0.3}$ \\ 
M2ccGLgp1 &  10  & 2 & no & yes & no & 72 & -261.7 & 0.1 & 2.1$^{+0.2}_{-0.2}$ & 1.1$^{+0.2}_{-0.2}$ & - \\ 
M4eeeeGCgp2 &  25  & 4 & no & yes & yes & 1.4 & -261.7 & 0.1 & 2.3$^{+0.2}_{-0.2}$ & 1.1$^{+0.3}_{-0.3}$ & 1.7$^{+0.3}_{-0.3}$ \\ 
M3cceGCgp1 &  16  & 3 & no & yes & yes & 72 & -261.8 & 0.0 & 2.2$^{+0.2}_{-0.2}$ & 1.0$^{+0.2}_{-0.2}$ & - \\ 
M2eeGLgp1 &  14  & 2 & no & yes & no & 72 & -261.9 & 0.1 & 2.1$^{+0.2}_{-0.2}$ & 1.1$^{+0.2}_{-0.2}$ & - \\ 
M3cccGCgp2 &  14  & 3 & no & yes & yes & 1.4 & -261.9 & 0.1 & 2.1$^{+0.2}_{-0.2}$ & 1.2$^{+0.3}_{-0.3}$ & 1.6$^{+0.3}_{-0.3}$ \\ 
M3eeeGCgp1 &  20  & 3 & no & yes & yes & 72 & -262.0 & 0.1 & 2.2$^{+0.2}_{-0.2}$ & 1.0$^{+0.2}_{-0.2}$ & - \\

\hline
\end{tabular}
\end{center}
\end{table*}

\begin{table*}
\tiny
\begin{center}
\caption{\label{ladder_2} Model ladder }
\begin{tabular}{cccccccccccc}
\hline
\hline
model name &  number of   & number of  & Jitter & Slope & Curvature & GP & LogZ & Bayes & semi-amplitude & semi-amplitude & semi-amplitude  \\
 &  parameter  & planets &  &  &  &  &  &  & planet b & planet c & planet d   \\
\hline
M3cceJCgp1$^{1.4}$ &  18  & 3 & yes & yes & yes & 72 & -262.1 & 0.1 & 2.0$^{+0.2}_{-0.2}$ & 1.1$^{+0.2}_{-0.2}$ & - \\ 
M3eeeJCgp1$^{1.4}$ &  22  & 3 & yes & yes & yes & 72 & -262.1 & 0.0 & 2.1$^{+0.2}_{-0.2}$ & 1.1$^{+0.2}_{-0.2}$ & - \\ 
M4eeeeJCgp2 &  27  & 4 & yes & yes & yes & 1.4 & -262.1 & 0.2 & 2.2$^{+0.2}_{-0.2}$ & 1.1$^{+0.3}_{-0.3}$ & 1.7$^{+0.3}_{-0.3}$ \\ 
M3eeeGCgp1$^{1.4}$ &  20  & 3 & no & yes & yes & 72 & -262.3 & 0.0 & 2.1$^{+0.1}_{-0.1}$ & 1.1$^{+0.1}_{-0.2}$ & - \\ 
M3eeeJCgp2 &  22  & 3 & yes & yes & yes & 1.4 & -262.3 & 0.1 & 2.2$^{+0.2}_{-0.3}$ & 1.1$^{+0.3}_{-0.3}$ & 1.7$^{+0.3}_{-0.3}$ \\ 
M2ccJCgp1 &  13  & 2 & yes & yes & yes & 72 & -262.4 & 0.3 & 2.0$^{+0.2}_{-0.2}$ & 1.0$^{+0.2}_{-0.2}$ & - \\ 
M3eeeGCgp2 &  20  & 3 & no & yes & yes & 1.4 & -262.7 & 0.1 & 2.3$^{+0.2}_{-0.2}$ & 1.2$^{+0.3}_{-0.4}$ & 1.6$^{+0.3}_{-0.3}$ \\ 
M2eeJCgp1 &  17  & 2 & yes & yes & yes & 72 & -262.8 & 0.6 & 2.1$^{+0.2}_{-0.2}$ & 1.1$^{+0.3}_{-0.3}$ & - \\ 
M3cccJCgp1$^{1.4}$ &  16  & 3 & yes & yes & yes & 72 & -263.4 & 0.9 & 2.0$^{+0.2}_{-0.2}$ & 1.0$^{+0.2}_{-0.2}$ & - \\ 
M4ccecGCgp1$^{1.4}$ &  19  & 4 & no & yes & yes & 72 & -264.3 & 0.1 & 2.0$^{+0.1}_{-0.1}$ & 1.2$^{+0.2}_{-0.2}$ & 1.2$^{+0.3}_{-0.4}$ \\ 
M4ccccGCgp1 &  17  & 4 & no & yes & yes & 72 & -264.4 & 1.3 & 2.2$^{+0.2}_{-0.2}$ & 1.1$^{+0.2}_{-0.2}$ & 1.3$^{+0.4}_{-0.4}$ \\ 
M3cceGCgp1$^{1.4}$ &  16  & 3 & no & yes & yes & 72 & -265.6 & 0.2 & 2.0$^{+0.2}_{-0.2}$ & 1.2$^{+0.2}_{-0.2}$ & - \\ 
M3cccGCgp1 &  14  & 3 & no & yes & yes & 72 & -265.8 & 3.0 & 2.2$^{+0.2}_{-0.2}$ & 1.1$^{+0.2}_{-0.2}$ & - \\ 
M4ccccGCgp1$^{1.4}$ &  17  & 4 & no & yes & yes & 72 & -268.9 & 0.1 & 2.1$^{+0.1}_{-0.1}$ & 1.0$^{+0.2}_{-0.2}$ & 1.3$^{+0.4}_{-0.4}$ \\ 
M1eJCgp1 &  12  & 1 & yes & yes & yes & 72 & -269.0 & 0.1 & 2.0$^{+0.3}_{-0.3}$ & - & - \\ 
M1cJCgp1 &  10  & 1 & yes & yes & yes & 72 & -269.1 & 0.8 & 1.9$^{+0.2}_{-0.2}$ & - & - \\ 
M2ccGCgp2 &  11  & 2 & no & yes & yes & 1.4 & -269.9 & 0.1 & 2.1$^{+0.2}_{-0.2}$ & 1.2$^{+0.3}_{-0.3}$ & - \\ 
M2ccJCgp2 &  13  & 2 & yes & yes & yes & 1.4 & -270.0 & 0.1 & 2.1$^{+0.2}_{-0.2}$ & 1.2$^{+0.3}_{-0.3}$ & - \\ 
M3cccGCgp1$^{1.4}$ &  14  & 3 & no & yes & yes & 72 & -270.1 & 0.1 & 2.1$^{+0.1}_{-0.1}$ & 1.0$^{+0.2}_{-0.2}$ & - \\ 
M2eeGCgp2 &  15  & 2 & no & yes & yes & 1.4 & -270.2 & 0.0 & 2.2$^{+0.2}_{-0.2}$ & 1.2$^{+0.4}_{-0.8}$ & - \\ 
M2eeJCgp2 &  17  & 2 & yes & yes & yes & 1.4 & -270.3 & 3.4 & 2.2$^{+0.3}_{-0.3}$ & 1.2$^{+0.4}_{-0.6}$ & - \\ 
M2eeGCgp1 &  15  & 2 & no & yes & yes & 72 & -273.7 & 0.1 & 2.1$^{+0.2}_{-0.2}$ & 1.1$^{+0.2}_{-0.2}$ & - \\ 
M2ccGCgp1 &  11  & 2 & no & yes & yes & 72 & -273.8 & 5.2 & 2.1$^{+0.2}_{-0.2}$ & 1.1$^{+0.2}_{-0.2}$ & - \\ 
M1eGLgp1 &  9  & 1 & no & yes & no & 72 & -279.0 & 2.8 & 2.1$^{+0.2}_{-0.2}$ & - & - \\ 
M1cGLgp1 &  7  & 1 & no & yes & no & 72 & -281.8 & 9.0 & 2.0$^{+0.2}_{-0.2}$ & - & - \\ 
M1eGCgp1 &  10  & 1 & no & yes & yes & 72 & -290.8 & 3.0 & 2.0$^{+0.2}_{-0.2}$ & - & - \\ 
M1cGCgp1 &  8  & 1 & no & yes & yes & 72 & -293.8 & 18.5 & 1.9$^{+0.2}_{-0.2}$ & - & - \\ 
M3eeeJC &  21  & 3 & yes & yes & yes & no & -312.3 & 0.4 & 2.2$^{+0.3}_{-0.3}$ & 0.9$^{+0.3}_{-0.3}$ & - \\ 
M3cceJC &  17  & 3 & yes & yes & yes & no & -312.6 & 5.7 & 1.7$^{+0.2}_{-0.2}$ & 1.3$^{+0.3}_{-0.2}$ & - \\ 
M3cccJC &  15  & 3 & yes & yes & yes & no & -318.3 & 0.4 & 1.7$^{+0.2}_{-0.2}$ & 1.2$^{+0.3}_{-0.3}$ & - \\ 
M2eeJC &  16  & 2 & yes & yes & yes & no & -318.7 & 0.5 & 1.9$^{+0.3}_{-0.3}$ & 1.2$^{+0.3}_{-0.3}$ & - \\ 
M2ccJC &  12  & 2 & yes & yes & yes & no & -319.2 & 1.1 & 1.7$^{+0.2}_{-0.2}$ & 1.1$^{+0.2}_{-0.3}$ & - \\ 
M3cceJgp1 &  16  & 3 & yes & no & no & 72 & -320.3 & 1.7 & 2.2$^{+0.2}_{-0.2}$ & 1.0$^{+0.2}_{-0.2}$ & - \\ 
M3cccJgp1 &  14  & 3 & yes & no & no & 72 & -322.1 & 0.0 & 2.2$^{+0.2}_{-0.2}$ & 1.1$^{+0.2}_{-0.2}$ & - \\ 
M3eeeJgp1 &  20  & 3 & yes & no & no & 72 & -322.1 & 1.4 & 2.2$^{+0.2}_{-0.2}$ & 1.0$^{+0.2}_{-0.2}$ & - \\ 
M2ccJgp1 &  11  & 2 & yes & no & no & 72 & -323.4 & 1.5 & 2.1$^{+0.2}_{-0.2}$ & 1.1$^{+0.2}_{-0.2}$ & - \\ 
M2eeJgp1 &  15  & 2 & yes & no & no & 72 & -324.9 & 1.8 & 2.1$^{+0.2}_{-0.2}$ & 1.1$^{+0.3}_{-0.3}$ & - \\ 
M1eJC &  11  & 1 & yes & yes & yes & no & -326.7 & 0.6 & 1.7$^{+0.3}_{-0.3}$ & - & - \\ 
M1cJC &  9  & 1 & yes & yes & yes & no & -327.4 & 2.8 & 1.6$^{+0.2}_{-0.2}$ & - & - \\ 
M1cJgp1 &  8  & 1 & yes & no & no & 72 & -330.2 & 0.3 & 2.0$^{+0.3}_{-0.3}$ & - & - \\ 
M1eJgp1 &  10  & 1 & yes & no & no & 72 & -330.5 & 1.4 & 2.0$^{+0.3}_{-0.3}$ & - & - \\ 
M3cceGgp1 &  14  & 3 & no & no & no & 72 & -331.9 & 1.5 & 2.2$^{+0.2}_{-0.2}$ & 1.0$^{+0.2}_{-0.2}$ & - \\ 
M3eeeGgp1 &  18  & 3 & no & no & no & 72 & -333.3 & 2.6 & 2.2$^{+0.2}_{-0.2}$ & 1.0$^{+0.2}_{-0.2}$ & - \\ 
M3cccGgp1 &  12  & 3 & no & no & no & 72 & -335.9 & 7.4 & 2.3$^{+0.2}_{-0.2}$ & 1.1$^{+0.2}_{-0.2}$ & - \\ 
M2ccGgp1 &  9  & 2 & no & no & no & 72 & -343.3 & 0.8 & 2.1$^{+0.2}_{-0.2}$ & 1.1$^{+0.2}_{-0.2}$ & - \\ 
M2eeGgp1 &  13  & 2 & no & no & no & 72 & -344.1 & 6.1 & 2.1$^{+0.2}_{-0.2}$ & 1.1$^{+0.2}_{-0.2}$ & - \\ 
M4eeeeJgp2 &  25  & 4 & yes & no & no & 1.4 & -350.3 & 7.5 & 2.5$^{+0.2}_{-0.2}$ & 1.5$^{+0.3}_{-0.3}$ & 2.2$^{+0.4}_{-0.4}$ \\ 
M4eeeeGgp2 &  23  & 4 & no & no & no & 1.4 & -357.7 & 2.9 & 2.5$^{+0.3}_{-0.3}$ & 1.6$^{+0.4}_{-0.4}$ & 2.3$^{+0.5}_{-0.4}$ \\ 
M1eGgp1 &  8  & 1 & no & no & no & 72 & -360.7 & 1.7 & 2.1$^{+0.2}_{-0.2}$ & - & - \\ 

\hline
\end{tabular}
\end{center}
\end{table*}

\begin{table*}
\tiny
\begin{center}
\caption{\label{ladder_2} Model ladder }
\begin{tabular}{cccccccccccc}
\hline
\hline
model name &  number of   & number of  & Jitter & Slope & Curvature & GP & LogZ & Bayes & semi-amplitude & semi-amplitude & semi-amplitude  \\
 &  parameter  & planets &  &  &  &  &  &  & planet b & planet c & planet d   \\
\hline

M3cccJgp2 &  14  & 3 & yes & no & no & 1.4 & -362.4 & 0.5 & 2.2$^{+0.3}_{-0.3}$ & 1.4$^{+0.4}_{-0.4}$ & 1.8$^{+0.5}_{-0.5}$ \\ 
M3eeeJL &  20  & 3 & yes & yes & no & no & -362.9 & 0.7 & 2.2$^{+0.4}_{-0.4}$ & 1.5$^{+0.3}_{-0.3}$ & - \\ 
M1cGgp1 &  6  & 1 & no & no & no & 72 & -363.6 & 0.1 & 2.0$^{+0.2}_{-0.2}$ & - & - \\ 
M4ccccJgp2 &  17  & 4 & yes & no & no & 1.4 & -363.7 & 0.5 & 2.2$^{+0.3}_{-0.3}$ & 1.5$^{+0.4}_{-0.4}$ & 1.8$^{+0.5}_{-0.5}$ \\ 
M3eeeJgp2 &  20  & 3 & yes & no & no & 1.4 & -364.2 & 2.3 & 2.3$^{+0.3}_{-0.3}$ & 1.4$^{+0.4}_{-0.4}$ & 1.9$^{+0.6}_{-0.5}$ \\ 
M2ccJgp2 &  11  & 2 & yes & no & no & 1.4 & -366.5 & 1.6 & 2.2$^{+0.3}_{-0.3}$ & 1.4$^{+0.4}_{-0.4}$ & - \\ 
M2eeJgp2 &  15  & 2 & yes & no & no & 1.4 & -368.1 & 0.0 & 2.2$^{+0.3}_{-0.3}$ & 1.4$^{+0.5}_{-0.6}$ & - \\ 
M3cceJL &  16  & 3 & yes & yes & no & no & -368.2 & 16.1 & 1.6$^{+0.2}_{-0.2}$ & 1.3$^{+0.3}_{-0.3}$ & - \\ 
M3cccGgp2 &  12  & 3 & no & no & no & 1.4 & -384.3 & 1.1 & 2.2$^{+0.3}_{-0.3}$ & 1.5$^{+0.4}_{-0.4}$ & 1.9$^{+0.5}_{-0.5}$ \\ 
M3eeeGgp2 &  18  & 3 & no & no & no & 1.4 & -385.4 & 0.1 & 2.3$^{+0.3}_{-0.3}$ & 1.4$^{+0.4}_{-0.4}$ & 2.2$^{+0.6}_{-0.5}$ \\ 
M4ccccGgp2 &  15  & 4 & no & no & no & 1.4 & -385.5 & 1.1 & 2.2$^{+0.3}_{-0.3}$ & 1.5$^{+0.4}_{-0.4}$ & 1.9$^{+0.5}_{-0.4}$ \\ 
M3cccJL &  14  & 3 & yes & yes & no & no & -386.5 & 2.6 & 1.9$^{+0.2}_{-0.2}$ & 1.6$^{+0.2}_{-0.3}$ & - \\ 
M2ccGgp2 &  9  & 2 & no & no & no & 1.4 & -389.2 & 0.8 & 2.2$^{+0.3}_{-0.3}$ & 1.4$^{+0.4}_{-0.4}$ & - \\ 
M2eeJL &  15  & 2 & yes & yes & no & no & -389.9 & 0.7 & 2.0$^{+0.3}_{-0.3}$ & 1.9$^{+0.3}_{-0.3}$ & - \\ 
M2eeGgp2 &  13  & 2 & no & no & no & 1.4 & -390.7 & 1.3 & 2.2$^{+0.3}_{-0.3}$ & 1.4$^{+0.5}_{-0.6}$ & - \\ 
M2ccJL &  11  & 2 & yes & yes & no & no & -392.0 & 15.1 & 1.8$^{+0.2}_{-0.2}$ & 1.5$^{+0.3}_{-0.3}$ & - \\ 
M1cJL &  8  & 1 & yes & yes & no & no & -407.1 & 0.4 & 1.6$^{+0.2}_{-0.2}$ & - & - \\ 
M1eJL &  10  & 1 & yes & yes & no & no & -407.5 & 144.9 & 1.7$^{+0.3}_{-0.3}$ & - & - \\ 
M3eeeGC &  19  & 3 & no & yes & yes & no & -552.4 & 19.6 & 1.8$^{+0.1}_{-0.1}$ & 2.5$^{+0.8}_{-1.2}$ & - \\ 
M3cceGC &  15  & 3 & no & yes & yes & no & -572.1 & 72.3 & 1.6$^{+0.1}_{-0.1}$ & 0.5$^{+0.1}_{-0.1}$ & - \\ 
M3cccGC &  13  & 3 & no & yes & yes & no & -644.4 & 7.3 & 1.6$^{+0.1}_{-0.1}$ & 1.4$^{+0.1}_{-0.1}$ & - \\ 
M3eeeJ &  19  & 3 & yes & no & no & no & -651.7 & 3.7 & 6.1$^{+1.1}_{-1.0}$ & 2.3$^{+0.4}_{-0.4}$ & - \\ 
M3cceJ &  15  & 3 & yes & no & no & no & -655.4 & 2.7 & 1.8$^{+0.2}_{-0.2}$ & 1.4$^{+0.2}_{-0.2}$ & - \\ 
M2eeGC &  14  & 2 & no & yes & yes & no & -658.1 & 8.8 & 1.8$^{+0.1}_{-0.1}$ & 1.2$^{+0.1}_{-0.1}$ & - \\ 
M2ccGC &  10  & 2 & no & yes & yes & no & -666.9 & 36.6 & 1.7$^{+0.1}_{-0.1}$ & 1.1$^{+0.1}_{-0.1}$ & - \\ 
M3cccJ &  13  & 3 & yes & no & no & no & -703.5 & 4.8 & 1.8$^{+0.3}_{-0.3}$ & 1.5$^{+0.3}_{-0.3}$ & - \\ 
M1eGC &  9  & 1 & no & yes & yes & no & -708.3 & 0.0 & 1.6$^{+0.1}_{-0.1}$ & - & - \\ 
M2eeJ &  14  & 2 & yes & no & no & no & -708.3 & 3.3 & 1.8$^{+0.1}_{-0.1}$ & 9.4$^{+0.5}_{-0.4}$ & - \\ 
M2ccJ &  10  & 2 & yes & no & no & no & -711.6 & 3.1 & 1.8$^{+0.2}_{-0.2}$ & 1.3$^{+0.3}_{-0.3}$ & - \\ 
M1cGC &  7  & 1 & no & yes & yes & no & -714.7 & 6.9 & 1.5$^{+0.1}_{-0.1}$ & - & - \\ 
M1cJ &  7  & 1 & yes & no & no & no & -721.6 & 0.1 & 1.7$^{+0.2}_{-0.2}$ & - & - \\ 
M1eJ &  9  & 1 & yes & no & no & no & -721.8 & 53.9 & 1.9$^{+0.3}_{-0.6}$ & - & - \\ 
M3eeeGL &  18  & 3 & no & yes & no & no & -775.7 & 68.3 & 2.0$^{+0.1}_{-0.1}$ & 5.4$^{+1.7}_{-2.5}$ & - \\ 
M3cceGL &  14  & 3 & no & yes & no & no & -844.0 & 150.9 & 1.5$^{+0.1}_{-0.1}$ & 0.7$^{+0.1}_{-0.1}$ & - \\ 
M3cccGL &  12  & 3 & no & yes & no & no & -994.9 & 14.7 & 1.8$^{+0.1}_{-0.1}$ & 1.5$^{+0.1}_{-0.1}$ & - \\ 
M2eeGL &  13  & 2 & no & yes & no & no & -1009.7 & 17.4 & 1.9$^{+0.1}_{-0.1}$ & 1.9$^{+0.1}_{-0.1}$ & - \\ 
M2ccGL &  9  & 2 & no & yes & no & no & -1027.0 & 81.5 & 1.7$^{+0.1}_{-0.1}$ & 1.5$^{+0.1}_{-0.1}$ & - \\ 
M1eGL &  8  & 1 & no & yes & no & no & -1108.6 & 0.7 & 1.5$^{+0.1}_{-0.1}$ & - & - \\ 
M1cGL &  6  & 1 & no & yes & no & no & -1109.3 & 769.2 & 1.5$^{+0.1}_{-0.1}$ & - & - \\ 
M3eeeG &  17  & 3 & no & no & no & no & -1878.5 & 156.2 & 9.9$^{+0.1}_{-0.1}$ & 9.9$^{+0.1}_{-0.1}$ & - \\ 
M3cceG &  13  & 3 & no & no & no & no & -2034.7 & 202.5 & 1.6$^{+0.1}_{-0.1}$ & 1.3$^{+0.1}_{-0.1}$ & - \\ 
M2eeG &  12  & 2 & no & no & no & no & -2237.3 & 56.2 & 9.8$^{+0.3}_{-0.1}$ & 2.3$^{+0.2}_{-0.1}$ & - \\ 
M3cccG &  11  & 3 & no & no & no & no & -2293.5 & 54.7 & 1.6$^{+0.1}_{-0.1}$ & 0.8$^{+0.1}_{-0.2}$ & - \\ 
M2ccG &  8  & 2 & no & no & no & no & -2348.2 & 0.3 & 1.6$^{+0.1}_{-0.1}$ & 1.2$^{+0.1}_{-0.1}$ & - \\ 
M1eG &  7  & 1 & no & no & no & no & -2348.6 & 52.9 & 6.1$^{+0.6}_{-0.2}$ & - & - \\ 
M1cG &  5  & 1 & no & no & no & no & -2401.4 & 0.0 & 1.5$^{+0.1}_{-0.1}$ & - & - \\

\hline
\end{tabular}
\end{center}
\end{table*}

\clearpage
%
%
%
%
%
%
%


\end{appendix}


\end{document}